\documentclass[traditabstract]{aa}
\usepackage{epsfig}    
\usepackage{lscape}
\usepackage{natbib}
\bibpunct{(}{)}{;}{a}{}{,}    
\usepackage{graphics}
\usepackage{graphicx,graphics}
\usepackage{color} 
\usepackage{times}
\usepackage{amsmath}
\usepackage{amssymb}
\begin{document}

\title {Circumnuclear regions of different BPT types in star-forming MaNGA galaxies: AGN detectability}

\author {
         L.~S.~Pilyugin\inst{\ref{MAO},\ref{ARI}} \and 
         E.~K.~Grebel\inst{\ref{ARI}} \and
         I.~A.~Zinchenko\inst{\ref{MAO},\ref{ARI}} \and
         M.~A.~Lara-L\'{o}pez\inst{\ref{DARK}} \and
         Y.~A.~Nefedyev\inst{\ref{KGU}} \and
         V.~M.~Shulga\inst{\ref{IC},\ref{CFC},\ref{RIAN}}
         }
\institute{Main Astronomical Observatory, National Academy of Sciences of Ukraine, 27 Akademika Zabolotnoho St, 03680, Kiev, Ukraine \label{MAO} \and
Astronomisches Rechen-Institut, Zentrum f\"{u}r Astronomie der Universit\"{a}t Heidelberg, M\"{o}nchhofstr.\ 12--14, 69120 Heidelberg, Germany \label{ARI} \and
DARK, Niels Bohr Institute, University of Copenhagen, Lyngbyvej 2, Copenhagen DK-2100, Denmark \label{DARK} \and 
Kazan Federal University, 18 Kremlyovskaya St., 420008, Kazan, Russian Federation \label{KGU} \and 
The International Center of Future Science of the Jilin University, 2699 Qianjin St., 130012, Changchun City, China \label{IC} \and
The International Center of Future Science, College of Physics, Jilin University, 2699 Qianjin St., 130012, Changchun City, China\label{CFC} \and 
Institut of Radio Astronomy of National Academy of Sciences of Ukraine, 4 Mystetstv str., 61002 Kharkov, Ukraine  \label{RIAN}
}

\abstract{
We consider the circumnuclear regions of star-forming MaNGA galaxies.
The spaxels spectra are classified as active-galactic-nucleus-like (AGN-like),  H\,{\sc ii}-region-like (or SF-like), and 
intermediate (INT) spectra  according to their positions on the Baldwin-Phillips-Terlevich (BPT) diagram. 
There are the following four configurations of the radiation distributions in the  circumnuclear regions in
(massive) galaxies:
1) AGN+INT, the innermost region of the AGN-like radiation is surrounded by a ring of  radiation
of the intermediate type;
2) INT, the central area of radiation of the intermediate type; 
3) SF+INT, the inner region of the   H\,{\sc ii}-region-like radiation is surrounded by a ring of  
radiation of the intermediate type; and 
4) SF, the central area of the H\,{\sc ii}-region-like radiation only.
The low ionization nuclear emission line regions (LINERs) of configurations 1 and 2 are examined.  
The spaxel spectra of the LINERs form a sequences on the BPT diagram, that is, they lie along the known
AGN-SF mixing line trajectories. The diagnostic line ratios of the spaxels spectra change smoothly with radius, 
from AGN-like (or INT) line ratios  at the galactic center to 
H\,{\sc ii}-region-like  at larger galactocentric distances. 
This is in agreement with the paradigm that the LINERs are excited by AGN activity. 
We found that the AGN and INT radiation in the circumnuclear region is accompanied by an enhanced 
gas velocity dispersion $\sigma_{gas}$. The radius of the area of the AGN and INT radiation
is similar to the radius of the area with enhanced $\sigma_{gas}$, and the central $\sigma_{gas,c}$
correlates with the luminosity of the AGN+INT area. 
We assume that the gas velocity dispersion can serve as an indicator of the AGN activity.
An appreciable enhancement of  $\sigma_{gas,c}$ was also measured in the SF-type centers of massive galaxies.
The values of  $\sigma_{gas,c}$ for the SF-type centers partly overlap with those of the AGN-type centers.  
This suggests that the manifestation of the circumnuclear region as AGN or as SF on the BPT diagram
depends not only on the value of $\sigma_{gas,c}$ (the level of the AGN activity) but it is also governed
by an additional parameter(s).
We find that there is a demarcation line between the positions of the AGN-type and SF-type objects on
the central gas velocity dispersion -- central H$\alpha$ surface brightness  diagram, 
in the sense that an object with a given value of   $\sigma_{gas,c}$ is  
an AGN-type  only if the central H$\alpha$  surface brightness is lower than some value. 
}


\titlerunning{Circumnuclear regions in galaxies}
\authorrunning{Pilyugin et al.}
\maketitle

\section{Introduction}

It has been known for a long time that active galactic nuclei (AGNs) are powered 
by accretion onto supermassive black holes \citep{Zeldovich1964,Salpeter1964,LyndenBell1969}.
It is believed \citep[e.g.,][]{Heckman2014} that the nuclei of most, quite possibly all, massive galaxies 
host a central supermassive black hole.
\citet{Heckman1980} suggested dividing the AGNs into two classes:  Low Ionization Nuclear Emission line Regions (LINERs), 
and Seyfert galaxies (types I and II), which are high ionization AGNs. 
The demarcation line between LINERs and Seyferts in  different diagrams are discussed in many works 
\citep[e.g.,][]{Kewley2006, CidFernandes2010}. 

It is believed at present that Seyfert and LINER phenomena are not fundamentally 
different in the sense that  photoionization is responsible for the emission of both classes of AGNs.
It is commonly accepted that the source of the ionizing photons in the Seyferts is the radiation of the 
accretion disk \citep{Shakura1973,Antonucci1993}.
On the contrary, the following different sources of the ionizing photons in the LINERs are discussed: the radiation of 
low-luminosity accretion disk \citep{Ferland1983}, the hot, low-mass evolved (postasymptotic giant branch) stars (HOLMES) 
\citep{Stasinska2006,Stasinska2008,Sarzi2010,YanBlanton2012,Singh2013}, and 
shocks caused by different phenomena (jets, galactic winds, galaxy-galaxy interactions)
\citep{Heckman1980,DopitaSutherland1995,Ho2014,Molina2018}. 

The intensities of strong, easily measured lines can be used to separate different types of 
emission-line objects according to their main  excitation mechanism (i.e., starburst or AGN). 
The most widely used system of spectral classification of emission-line objects is the 
[O\,{\sc iii}]$\lambda$5007/H$\beta$ versus \ [N\,{\sc ii}]$\lambda$6584/H$\alpha$ diagnostic diagram 
suggested by \citet{Baldwin1981}  (BPT classification diagram). The excitation properties of objects 
with the emission line specta are studied by plotting the low-excitation [N\,{\sc ii}]$\lambda$6584/H$\alpha$ 
line ratio against the high-excitation [O\,{\sc iii}]$\lambda$5007/H$\beta$ line ratio. 
It was noted \citep{Stasinska2006,Stasinska2008} that the galaxies from the Sloan Digital Sky 
Survey \citep{York2000} occupy a well-defined region, evoking the wings of a flying seagull. The 
left wing consists of star-forming galaxies while the right wing is attributed to galaxies 
with active nucleus. 
The exact location of the dividing line between starbursts (H\,{\sc ii} regions)  
and AGNs is still controversial 
\citep[see, e.g.,][]{Kewley2001,Kauffmann2003,Stasinska2006,Stasinska2008,Herpich2016}.  

The AGNs are revealed in many galaxies using the BPT classification diagram
\citep[e.g.,][]{Ho1997,Kauffmann2003}.
At the same time, some AGNs are not detected by means of the BPT classification diagram
in a number of  massive galaxies. 
Indeed, some galaxies identified as AGNs by  their hard X-ray luminosity are 
not classified as AGN by their optical emission line properties 
\citep{Elvis1981,Moran2002,Pons2014,Pons2016,Agostino2019}. 
This suggests that the presence of the AGN in the galaxy  does not necessarily result
in the appearance of  the LINER (or Seyfert) on the BPT diagram.  

The Integral Field Unit (IFU) spectroscopy reveals that many galaxies involve 
low-ionization emission-line regions (LIER) located far from the centers of galaxies 
\citep[e.g.,][]{Belfiore2016,Hviding2018,Parkash2019}.
This evidences that LIERs are not produced by  AGNs, 
instead, the source of the ionizing photons in the LIERs can be attributed to  
hot low-mass evolved stars (HOLMES).  
 
Since {\it i)} the presence of the AGN is not necessarily accompanied by the
LINER (or Seyfert) on the BPT diagram  and {\it ii)} the LIERs are excited by the HOLMES but not the AGNs 
then one can suggest that the LINERs (at least some of them) 
be also attributed to the HOLMES rather than to the AGNs, this is,  
LINERs differ from LIERs by their locations in galaxies only. 
Or they differ also by the sources of the ionizing photons. 
Here we examine the properties of the circumnuclear
regions in the MaNGA galaxies with and without detected AGNs aiming to clarify 
the sources of the ionizing photons in the LINERs and why the
AGNs are not detected in some massive galaxies through the BPT classification diagram. 

This paper is organized in the following way. The data are described in
Section 2. In Section 3 the properties of the circumnuclear
regions in the MaNGA galaxies are examined. The discussion is
given in  Section 4. Section 5 contains a brief summary.

\section{Data}

\begin{figure}
\begin{center}
\resizebox{1.000\hsize}{!}{\includegraphics[angle=000]{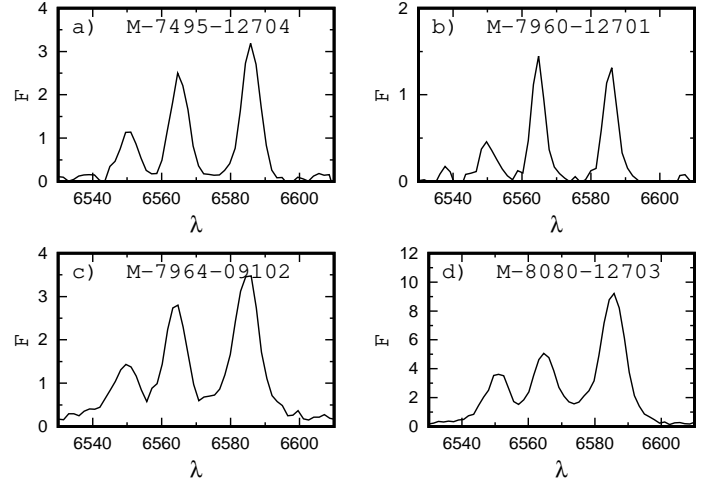}}
\caption{
Spectra of the central regions in four MaNGA galaxies.
The flux $F$ is in units of 10$^{-17}$ erg/s/cm$^2$, and the wavelength $\lambda$ in angstroms.
The stacked spectra of nine central spaxels are shown in each panel.
}
\label{figure:spectraexamples}
\end{center}
\end{figure}

\begin{figure}
\begin{center}
\resizebox{1.000\hsize}{!}{\includegraphics[angle=000]{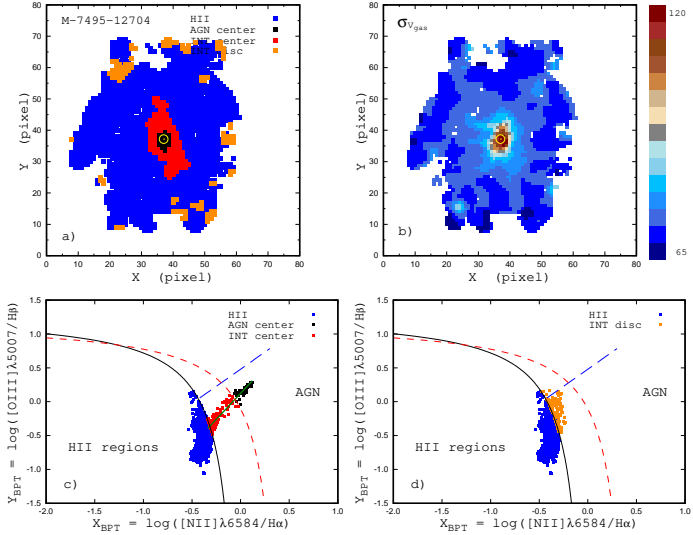}}
\caption{
Example of BPT diagrams for a galaxy with the circumnuclear region classified as AGN type.
{\em Panel} $a$: locations of the spaxels with the AGN-like,  H\,{\sc ii}-region-like,
and intermediate spectra on the image of  the MaNGA galaxy M-7495-12704 in sky coordinates
(pixels, the physical scale is 0.309 kpc/pixel).
The BPT type of the spectra is color-coded.
The spaxels with the intermediate spectra in the circumnuclear region (center) and
in the disc are shown by different colors.
The circle marks the kinematic center of the galaxy.
{\em Panel} $b$: map of the observed (non corrected for instrumental profile)
gas velocity dispersion. The value of gas velocity dispersion is color-coded.  
{\em Panel} $c$:  BPT diagram for the individual spaxels with AGN (black
symbols) and intermediate (red symbols) spectra classification in the circumnuclear region. 
The thick solid line is the Y$_{BPT}$ = $a$X$_{BPT}$ + $b$ relation obtained for those data.
Solid and short-dashed curves mark the demarcation line between AGNs and H\,{\sc ii}
regions defined by \citet{Kauffmann2003} and \citet{Kewley2001}, respectively.
The long-dashed line is the dividing line between Seyfert galaxies and
LINERs defined by \citet{CidFernandes2010}.
The blue points are the  H\,{\sc ii}-region-like spectra in the galaxy.
{\em Panel} $d$ shows the same as {\em panel} $b$ but for the spaxels with the  
intermediate-type spectra located in the disc.
}
\label{figure:bpt-definition}
\end{center}
\end{figure}

\begin{figure*}
\begin{center}
\resizebox{1.00\hsize}{!}{\includegraphics[angle=000]{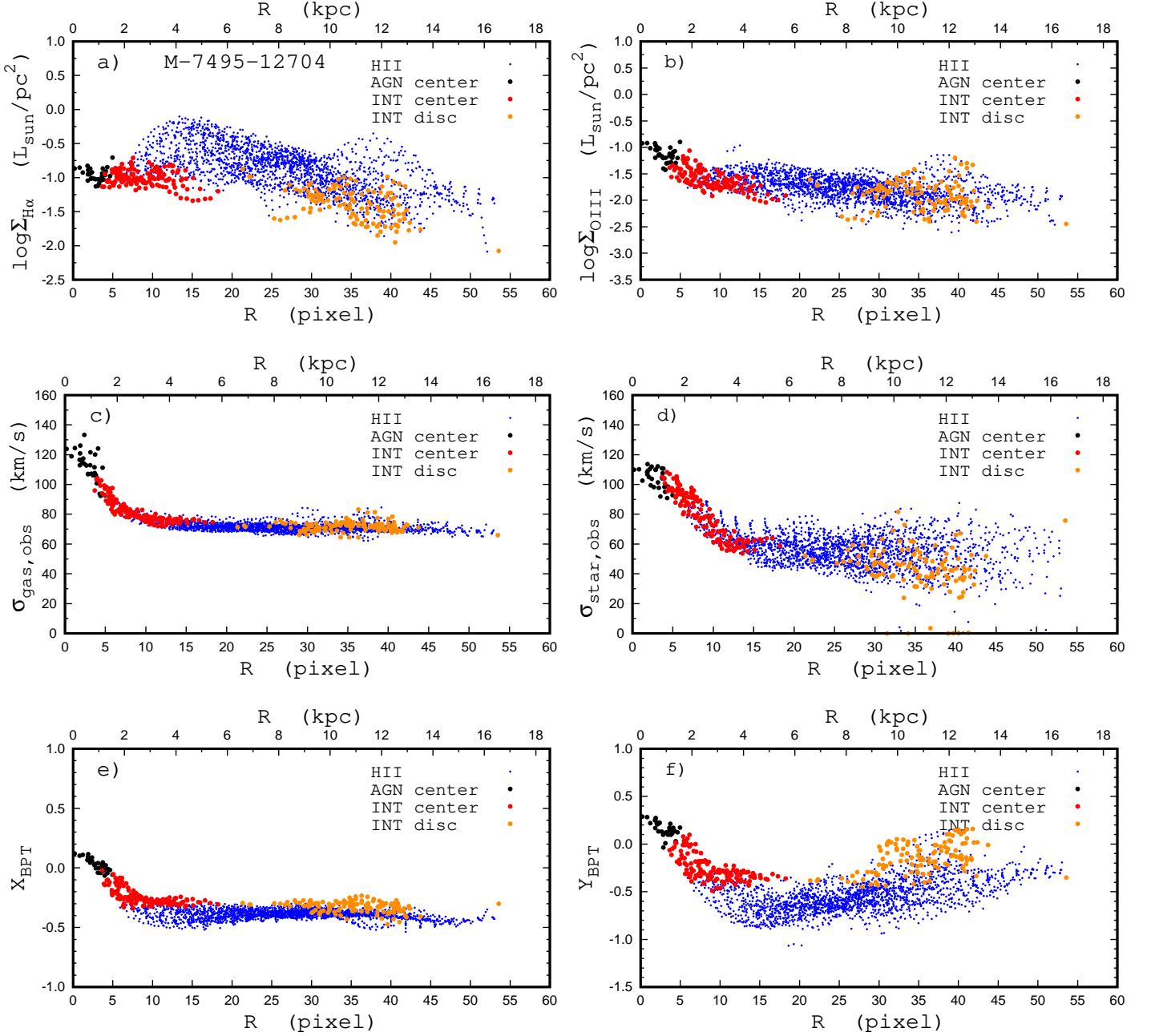}}
\caption{
Radial variation of properties in the disc
of the MaNGA galaxy M-7495-12704 with the circumnuclear region of the
AGN type (Fig.~\ref{figure:bpt-definition}).
{\em Panels} show the surface brightness in  H$\alpha$ ({\em panel} $a$)  
and  [O\,{\sc iii}]$\lambda\lambda$4959,5007 ({\em panel} $b$) lines,
the observed (non corrected for instrumental profile) velocity dispersion of the gas $\sigma_{gas,obs}$ ({\em panel} $c$)  
and stars $\sigma_{star,obs}$ ({\em panel} $d$),
the coordinates in the BPT diagram X$_{BPT}$ = log([N\,{\sc ii}]$\lambda$6584/H$\alpha$) ({\em panel} $e$), 
and Y$_{BPT}$ = log([O\,{\sc iii}]$\lambda$5007/H$\beta$) ({\em panel} $f$),
as a function of the galactocentric distance. 
The BPT type of the spectra is color-coded.
The spaxels classified as intermediate-type  in the disc and in
the circumnuclear region (center) are shown by different colors.
}
\label{figure:rg-param-example}
\end{center}
\end{figure*}

\begin{figure}
\begin{center}
\resizebox{1.00\hsize}{!}{\includegraphics[angle=000]{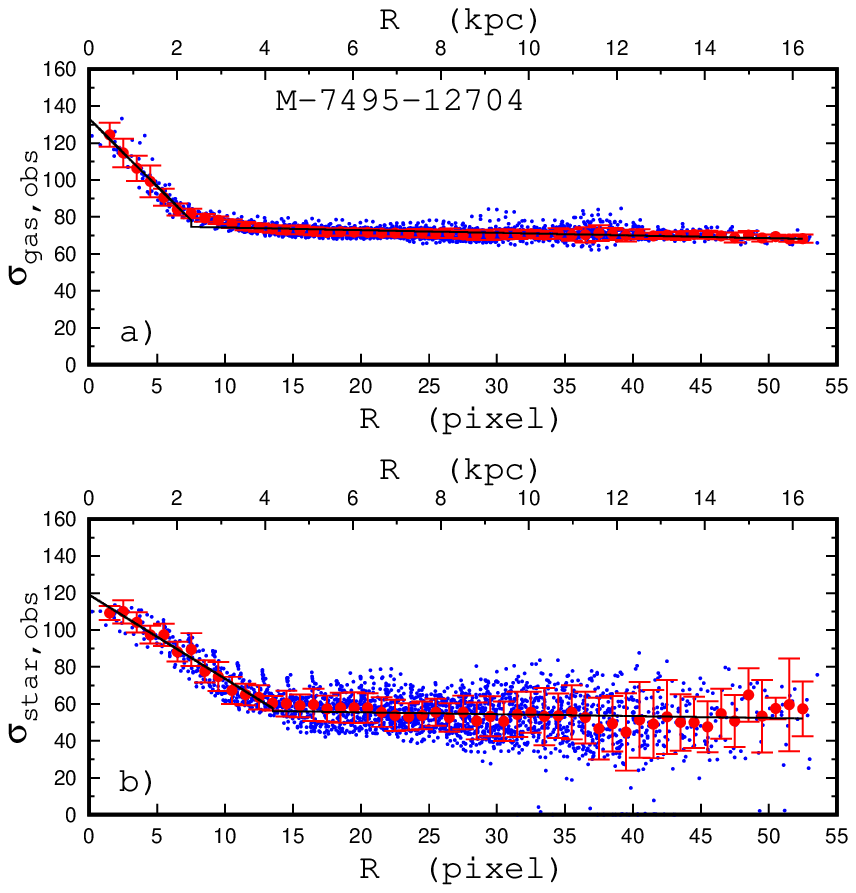}}
\caption{
Determination of the radii of the zones of the enhanced gas and stellar velocity dispersions.
{\em Panel} $a$ shows the  observed (non corrected for the instrumental
profile) gas velocity dispersion $\sigma_{gas,obs}$ as a function of radius 
for individual spaxels (blue points) and median values of $\sigma_{gas,obs}$ 
with the mean deviations (red points and bars) for intervals corresponding to the width of 1 spaxel.
The broken line is the fit to the median values of $\sigma_{gas,obs}$.
{\em Panel} $b$ shows the same as panes $a$ but for the 
observed stellar velocity dispersion $\sigma_{star,obs}$. 
}
\label{figure:r}
\end{center}
\end{figure}

The spectroscopic measurements from the SDSS-IV MaNGA survey
\citep{Bundy2015,Albareti2017} provide the possibility to measure the
surface brightness distribution and to determine photometric characteristics 
(such as the optical radius and luminosity), to measure the line-of-sight 
velocity field and to derive kinematic characteristics (kinematic angles 
and rotation cuerve), and to measure emission line fluxes and obtain 
abundance maps. 
The spectrum of each spaxel is reduced as indicated in
\citet{Zinchenko2016}.  The stellar background is fitted using
the public version of the STARLIGHT code
\citep{CidFernandes2005,Mateus2006,Asari2007} adapted for execution in
the NorduGrid ARC\footnote{http://www.nordugrid.org/} environment of
the Ukrainian National Grid.  
The library of 150 synthetic simple stellar population (SSP) spectra from the
evolutionary synthesis models by \citet{Bruzual2003} with ages from
1~Myr up to 13~Gyr and metallicities of $Z = 0.0001$, 0.004, 0.008,
0.02, and 0.05 are used. The reddening law of \citet{Cardelli1989}
with $R_V = 3.1$ is adopted.  The obtained stellar radiation contribution was
subtracted from the observed spectrum. 

The  emission lines in each spaxel spectrum were estimated 
using a Gaussian fit to the line profiles. The emission line parameters 
include the central wavelength $\lambda_{0}$, the sigma $\sigma$, and the flux $F$. 
For each spectrum, we measure the 
[O\,{\sc ii}]$\lambda\lambda$3727,3729,
H$\beta$,  
[O\,{\sc iii}]$\lambda$5007,
H$\alpha$, and   
[N\,{\sc ii}]$\lambda$6584 emission lines. 
The measured fluxes in those lines are used for the interstellar reddening correction, 
the determination of the BPT type of the spectrum (separation of the 
emission-line objects according to their main excitation mechanism), 
and for the abundance determination. The central wavelength of the 
H$\alpha$ line is converted into the line-of-sight velocity. 
The sigma of the best-fit Gaussin $\sigma_{{\rm H}\alpha}$ specifies 
the width of the  H$\alpha$ line
(the full width at half maximum FWHM and $\sigma$ is linked by
the relation FWHM $\approx$ 2.355 $\sigma$)  
and is converted into the observed  
gas velocity dispersion $\sigma_{gas,obs}$ as 
$\sigma_{gas,obs}$ =  ($\sigma_{{\rm H}\alpha}$/$\lambda_{{\rm H}\alpha}$)$c$,
where  $c$ is the speed of light. 
The observed gas velocity dispersion $\sigma_{gas,obs}$ in the spaxels within 
the circumnuclear region is corrected for the instrumental profile 
in the following way $\sigma_{gas}^{2}$ = $\sigma_{gas,obs}^{2}$ -- $\sigma_{median,{\rm HII}}^{2}$, 
where $\sigma_{median,{\rm HII}}$ is the median value of the $\sigma_{gas,obs}$ for 
all the spaxels with the  H\,{\sc ii}-region-like spectra in the galaxy 
(see discussion below).
   
It should be noted that only those spaxel spectra where all the used lines 
were measured with a signal-to-noise ratio S/N $> 3$ were considered.  
The measured emission line fluxes were corrected for interstellar reddening
using the theoretical H$\alpha$/H$\beta$ ratio and the reddening
function from \citet{Cardelli1989} for $R_{V}$ = 3.1.  We assume
$C_{{\rm H}{\beta}} = 0.47A_{V}$ \citep{Lee2005}.
 
The oxygen abundances were obtained through the three-dimensional $R$ calibration \citep{Pilyugin2016,Pilyugin2018}.
The [O\,{\sc ii}]$\lambda\lambda$3727,3729, H$\beta$, [O\,{\sc iii}]$\lambda\lambda$4959,5007, 
and [N\,{\sc ii}]$\lambda\lambda$6548,6584 emission lines are used for oxygen abundance determinations. 
The flux in the [O\,{\sc iii}]$\lambda\lambda$4959,5007 lines is estimated as
[O\,{\sc iii}]$\lambda\lambda$4959,5007 = 1.3[O\,{\sc iii}]$\lambda$5007
and the flux in the [N\,{\sc ii}]$\lambda\lambda$6548,6584 lines is estimated as 
[N\,{\sc ii}]$\lambda,\lambda$6548,6584 = 1.3[N\,{\sc ii}]$\lambda$6584  \citep{Storey2000}.
The $R$-calibration produces abundances compartible to the $T_{e}$-based abundance scale and
is workable over the whole metallicity scale of H\,{\sc ii} regions. 

To determine the galactocentric distances of the spaxels,
the geometrical parameters of the galaxy are needed.
The parameters of the galaxy are determined from the observed velocity
field in the standard way assuming that a galaxy is  a
symmetrically rotating disc
\citep[e.g.,][]{Warner1973,Begeman1989,deBlok2008,Oh2018}.
The position of the kinematic center of the galaxy, the position 
angle of the major kinematic axis and the kinematic inclination angle 
are determined from the measured line-of-sight velocities of the H$\alpha$ line 
in the same way as in our previous papers \citep{Pilyugin2019,Pilyugin2020,Zinchenko2019}. 
In brief, the observed line-of-sight velocities recorded on a set of pixel
coordinates (1 pixel = 0.5 arcsec for MaNGA galaxies) are related
to the kinematical parameters of the galaxy and its
rotation curve. The deprojected galaxy plane is divided into rings with a width
of one pixel. The rotation velocity is assumed to be the same for all the points
within the ring. The position angle of the major axis and the galaxy inclination
angle are assumed to be the same for all the rings. The coordinates of the rotation
center of the galaxy, the position angle of the major axis,  the galaxy inclination
angle, and the rotation curve are derived through the best fit to the observed
velocity field. All the spaxels with measured  line-of-sight velocities are ased.

The distances to the galaxies were adopted from {\sc
ned}.  The {\sc ned} distances use flow corrections for Virgo, the
Great Attractor, and Shapley Supercluster infall (adopting a
cosmological model with $H_{0} = 73$ km/s/Mpc, $\Omega_{m} = 0.27$,
and $\Omega_{\Lambda} = 0.73$).

The SDSS data base provides values of the stellar masses of its target
galaxies determined in different ways.  We have chosen the
spectroscopic $M_{sp}$ masses of the SDSS and BOSS galaxies
\citep[BOSS stands for the Baryon Oscillation Spectroscopic Survey in
SDSS-III, see][]{Dawson2013}.  The spectroscopic masses, $M_{sp}$,
are the median (50th percentile of the probability distribution
function, PDF) of the logarithmic stellar masses from table {\sc
stellarMassPCAWiscBC03} determined by the Wisconsin method
\citep{Chen2012} with the stellar population synthesis models from
\citet{Bruzual2003}.

We selected a sample of MaNGA galaxies to be analyzed by considering 
the surface brightness maps, the line-of-sight gas velocity fields,  
the distribution of the spaxels of the AGN-like, H\,{\sc ii}-region-like, 
and intermediate spectra on the image of the MaNGA galaxies, and finally, the 
type of spectra at the center of the galaxies. 
First, we search for galaxies with intermediate and AGN-like spectra at the 
centers. The circumnuclear region in such galaxy is defined as the central region
involving the spaxels with AGN-like and intermediate spectra.  
The map of the distributions of the spaxels with the AGN-like, H\,{\sc ii}-region-like, 
and intermediate spectra on the images for around three thouthands MaNGA galaxies of 
the DR15 release were constructed. We search for galaxies with intermediate and AGN-like
spectra at the centers by visual inspection of those maps.
Then we selected galaxies using the following criteria.  \\

Only the star-forming galaxies (where emission lines are measured in
spectra of more than hundred spaxels)  were considered, since the characteristics 
to be analyzed (emission line ratios, gas velocity dispersion, luminosities
and surface brightnesses in the emission lines) are based on emission lines.  \\

Galaxies measured with a small number of fibers (19 and 37) were excluded. \\

In order to exclude a peculiar and strongly interacting or merging galaxies,
we required that the surface brightness distribution is centrally symmetric and smooth
and the line-of-sight gas velocity field are rather regular, this is, the isovelocity
curves of the measured velocity field are more or less close to a set of parabolic
curves (the hourglass-like picture for a rotating disc).   \\

We required that the spaxels with measured emission lines cover an area larger
than the circumnuclear region,
this is, the circumnuclear region is measured entirely and does not extend
beyond the measured area. 
We also required that the emission lines measurements were available for all (or at least for the 
majority of) the spaxels within the circumnuclear area. 
Under those conditions, the circumnuclear area (the central region with the AGN-like and intermediate spectra)  
can be outlined and its characteristics (e.g., luminosity) can be obtained. \\

We selected 85 MaNGA galaxies with those criteria.

The emission line profiles of the spectra at the centers of the majority of galaxies
with AGN-like and intermediate spectra selected with the above criteria
are rather simple (panels a and b of Fig.~\ref{figure:spectraexamples}) 
and can be fitted by single Gaussians. 
A broad component can be seen in the emission line profiles in the spaxel spectra
at the centers of some selected galaxies
(panels c and d of Fig.~\ref{figure:spectraexamples}).
Galaxies with a broad component in the emission line spectra at the center were excluded
from the current consideration.  The single Gaussian is not adequate approximation
for the profile of the emission line with a broad component, such lines should be fitted
by two Gaussians. The investigation of the properties of the AGNs with the broad
components in the emission lines should be the subject of the separate study. For the same reason,
the galaxies classified as Seyfert galaxies according to the BPT diagram were also excluded
from the current consideration.  \\

Our final list includes 46 galaxies with the circumnuclear region with  AGN-like radiation,
and 17 galaxies with the circumnuclear region of the intermediate type.
The selected galaxies are listed in Tables \ref{table:sampleAGN} and  \ref{table:sampleINT} .

For comparison purposes, we also consider 98 galaxies with the circumnuclear region with the 
H\,{\sc ii} region-like radiation. 
Those galaxies are selected using the same criteria as the galaxies with LINERs,
except with the requirement for H\,{\sc ii} region-like spectra at the center.

\begin{table*}
\caption[]{\label{table:sampleAGN}
Properties of our sample of MaNGA galaxies with the circumnuclear region classified as AGN-type.  
The columns show
the name (the MaNGA number),
the galaxy distance $d$ in Mpc,
the spectroscopic stellar mass $M_{sp}$ in solar masses,
the kinematic inclination angle $i$,
the radius of the zone of the dominant AGN conribution to the radiation $R_{AGN}$ in kpc,
the radius of the zone of the influence of the AGN on the radiation $R_{INT}$ in kpc,
the radius of the zone of the enhanced gas velocity dispersion $R_{\sigma_{gas}}$ in kpc,
the H$\alpha$ line luminosity of the zone with dominant AGN contribution $R_{AGN,{\rm H}\alpha}$, 
the H$\alpha$ line luminosity of the zone of AGN influence $R_{INT,{\rm H}\alpha}$, 
the central surface brightness in the  H$\alpha$ line $\Sigma_{{\rm H}\alpha,C}$ in $L_{\sun}/pc^{2}$,
the central gas velocity dispersion $\sigma_{gas,c}$ in km/s,
the median value of the observed (non corrected for the instrumental profile) gas velocity 
dispersions for all the spaxels with  H\,{\sc ii}-region-like spectra  $\sigma_{gas,HIIobs}$ in km/s. 
}
\begin{center}
\begin{tabular}{cccccccccccc} \hline \hline
Name                  &
d                     &
M$_{sp}$               &
$i$                   &
R$_{AGN}$              &
R$_{INT}$              &
R$_{\sigma_{gas}}$        &
logL$_{AGN,Ha}$        &
logL$_{INT,Ha}$        &           
log$\Sigma_{Ha,C}$      &           
log$\sigma_{gas,c}$     &
log$\sigma_{gas,HIIobs}$     \\
                      &
(Mpc)                 &
(M$_{\sun}$)           &
($\degr$)             &
(kpc)                &           
(kpc)                 &           
(kpc)                 &
(erg/s)               &
(erg/s)               &           
(L$_{\sun}/pc^{2}$)     &           
km/s                  &           
km/s             \\   \hline
 7495 12704 &  127.5 &   10.87 &  57.5 &    1.35 &    3.07 &    2.32 &  39.39 &  40.07 &  -0.902 &   2.019 &   1.852 \\
 7960 12701 &  136.8 &   10.97 &  30.4 &    0.94 &    2.12 &    2.16 &  39.21 &  39.83 &  -0.884 &   1.729 &   1.859 \\
 7968 06101 &  209.4 &   10.81 &  29.9 &    0.87 &    2.90 &    2.80 &  38.98 &  40.13 &  -0.929 &   1.852 &   1.864 \\
 8078 06103 &  113.2 &   10.95 &  53.3 &    0.98 &    2.02 &    2.33 &  39.87 &  40.40 &  -0.302 &   1.885 &   1.902 \\
 8083 12704 &   91.4 &   10.86 &  31.7 &    1.18 &    2.15 &    1.89 &  39.29 &  39.73 &  -0.763 &   1.874 &   1.839 \\
 8140 12703 &  133.3 &   11.22 &  52.5 &    2.12 &    3.92 &    3.07 &  40.28 &  40.52 &   0.014 &   2.141 &   1.855 \\
 8144 12703 &  160.0 &   11.37 &  24.4 &    0.79 &    1.64 &    1.75 &  39.40 &  40.04 &  -0.453 &   1.943 &   1.844 \\
 8146 12702 &  257.6 &   11.24 &  53.6 &    0.91 &    3.39 &    4.07 &  39.58 &  40.63 &  -0.425 &   2.086 &   1.867 \\
 8158 12702 &  307.1 &   11.31 &  36.9 &    1.69 &    3.48 &    3.36 &  39.84 &  40.44 &  -0.666 &   2.163 &   1.855 \\
 8243 12701 &  186.4 &   11.64 &  69.4 &    2.78 &    4.82 &    4.75 &  40.25 &  40.63 &  -0.482 &   2.038 &   1.876 \\
 8249 12704 &  114.1 &   11.13 &  34.4 &    1.71 &    3.03 &    2.91 &  39.45 &  39.95 &  -1.154 &   1.971 &   1.859 \\
 8257 06101 &  124.8 &   10.54 &  43.7 &    0.57 &    1.62 &    1.67 &  38.95 &  39.87 &  -0.623 &   1.964 &   1.872 \\
 8257 12705 &  153.6 &   10.93 &  70.2 &    1.25 &    3.25 &    3.91 &  39.79 &  40.43 &  -0.469 &   1.913 &   1.860 \\
 8313 09101 &  164.7 &   10.98 &  44.3 &    0.80 &    2.58 &    2.60 &  39.72 &  40.52 &  -0.225 &   2.106 &   1.868 \\
 8313 09102 &  143.2 &   10.92 &  43.4 &    2.11 &    3.68 &    2.61 &  39.66 &  40.19 &  -0.968 &   2.019 &   1.861 \\
 8318 12703 &  167.3 &   11.36 &  50.5 &    1.81 &    3.06 &    4.26 &  39.36 &  39.95 &  -1.227 &   2.027 &   1.854 \\
 8320 09102 &  222.9 &   11.31 &  48.0 &    2.44 &    4.35 &    3.52 &  39.77 &  40.30 &  -0.996 &   2.146 &   1.856 \\
 8326 09102 &  300.3 &   11.26 &  28.8 &    2.19 &    3.77 &    4.01 &  39.50 &  40.02 &  -1.288 &   2.101 &   1.827 \\
 8332 12705 &  143.2 &   11.13 &  41.4 &    1.57 &    2.68 &    2.26 &  39.79 &  40.27 &  -0.515 &   2.067 &   1.856 \\
 8444 06102 &  112.7 &   10.68 &  68.0 &    0.56 &    2.35 &    2.60 &  38.91 &  40.03 &  -0.667 &   1.850 &   1.885 \\
 8448 12703 &  308.4 &   11.18 &  64.7 &    2.33 &    4.91 &    4.12 &  39.67 &  40.26 &  -1.307 &   2.092 &   1.861 \\
 8484 12702 &  238.7 &   11.09 &  55.4 &    2.03 &    4.90 &    3.77 &  39.56 &  40.23 &  -1.098 &   2.046 &   1.851 \\
 8550 06103 &  108.6 &   10.67 &  40.8 &    1.05 &    1.65 &    1.45 &  39.73 &  40.09 &  -0.356 &   2.040 &   1.896 \\
 8550 12702 &  131.9 &   11.15 &  66.1 &    1.94 &    2.79 &    3.04 &  40.22 &  40.38 &  -0.202 &   2.004 &   1.910 \\
 8550 12705 &  130.1 &   11.34 &  54.8 &    1.86 &    2.76 &    2.68 &  39.80 &  40.20 &  -0.857 &   2.110 &   1.866 \\
 8591 06104 &  180.5 &   11.04 &  38.0 &    0.88 &    2.31 &    2.85 &  39.11 &  40.16 &  -0.871 &   1.933 &   1.861 \\
 8601 12705 &  129.1 &   10.80 &  66.7 &    1.56 &    3.68 &    3.91 &  39.45 &  40.10 &  -0.874 &   1.888 &   1.859 \\
 8612 12702 &  264.5 &   11.30 &  37.2 &    2.95 &    4.50 &    4.17 &  40.27 &  40.61 &  -0.636 &   2.040 &   1.842 \\
 8713 06102 &  295.4 &   11.24 &  38.5 &    2.54 &    5.09 &    3.95 &  39.83 &  40.51 &  -0.968 &   2.073 &   1.850 \\
 8714 06102 &  215.9 &   11.52 &  34.8 &    1.42 &    2.42 &    2.36 &  40.22 &  40.63 &  -0.205 &   2.034 &   1.860 \\
 8719 06103 &  247.3 &   11.46 &  33.6 &    1.57 &    3.40 &    2.70 &  40.02 &  40.77 &  -0.455 &   2.099 &   1.884 \\
 8977 06103 &  253.5 &   11.25 &  31.8 &    1.36 &    3.28 &    4.00 &  39.48 &  40.41 &  -0.860 &   1.818 &   1.874 \\
 8979 12701 &  325.2 &   11.36 &  50.6 &    3.49 &    6.56 &    5.92 &  39.74 &  40.48 &  -1.547 &   2.076 &   1.836 \\
 8984 12705 &  158.3 &   11.29 &  36.5 &    1.85 &    3.27 &    3.27 &  40.04 &  40.57 &  -0.543 &   2.173 &   1.860 \\
 8996 12705 &  199.8 &   11.14 &  59.5 &    2.57 &    4.85 &    4.12 &  39.54 &  40.35 &  -1.380 &   2.016 &   1.855 \\
 8997 12701 &  199.8 &   11.14 &  56.8 &    1.85 &    4.81 &    3.15 &  39.37 &  40.37 &  -1.248 &   2.009 &   1.858 \\
 8999 06103 &  278.9 &   11.55 &  68.6 &    2.89 &    5.58 &    6.43 &  40.47 &  40.93 &  -0.586 &   2.007 &   1.928 \\
 9028 12701 &  204.9 &   10.63 &  35.6 &    0.93 &    2.62 &    2.74 &  39.09 &  39.95 &  -0.916 &   1.972 &   1.851 \\
 9041 12703 &  237.8 &   10.87 &  60.8 &    2.13 &    4.16 &    3.75 &  39.94 &  40.50 &  -0.805 &   1.909 &   1.846 \\
 9042 09102 &  136.2 &   10.81 &  58.3 &    1.09 &    2.41 &    3.14 &  39.13 &  39.98 &  -0.992 &   1.891 &   1.866 \\
 9049 12702 &  378.0 &   11.33 &  49.7 &    2.87 &    5.82 &    6.88 &  40.44 &  40.89 &  -0.609 &   2.098 &   1.835 \\
 9183 12702 &  137.4 &   10.85 &  54.1 &    0.74 &    2.11 &    2.83 &  39.37 &  40.22 &  -0.401 &   1.871 &   1.892 \\
 9500 06103 &  270.2 &   11.42 &  17.3 &    0.76 &    2.42 &    3.61 &  39.61 &  40.57 &  -0.224 &   1.808 &   1.852 \\
 9505 12701 &  191.2 &   11.29 &  68.4 &    1.88 &    4.33 &    4.87 &  39.31 &  40.26 &  -1.230 &   2.008 &   1.888 \\
 9870 09101 &  157.3 &   11.08 &  35.2 &    2.21 &    3.18 &    2.10 &  40.20 &  40.42 &   0.001 &   2.146 &   1.863 \\
 9881 06102 &  119.2 &   11.00 &  20.5 &    0.81 &    1.36 &    1.30 &  39.82 &  40.29 &  -0.056 &   2.025 &   1.868 \\
\hline
\end{tabular}\\
\end{center}
\begin{flushleft}
\end{flushleft}
\end{table*}

\begin{table*}
\caption[]{\label{table:sampleINT}
Properties of our sample of MaNGA galaxies with the circumnuclear region classified as INT type.  
The columns show
the name (the MaNGA number),
the galaxy distance $d$ in Mpc,
the spectroscopic stellar mass $M_{sp}$ in solar masses,
the kinematic inclination angle $i$,
the radius of the zone of the influence of the AGN on the radiation $R_{INT}$ in kpc,
the radius of thezone of the enhanced gas velocity dispersion $R_{\sigma_{gas}}$ in kpc,
the H$\alpha$ line luminosity of the zone of AGN influence $R_{INT,{\rm H}\alpha}$ 
the central surface brightness in the  H$\alpha$ line $\Sigma_{{\rm H}\alpha,C}$ in $L_{\sun}/pc^{2}$,
the central gas velocity dispersion $\sigma_{gas,c}$ in km/s,
the median value of the observed (non corrected for the instrumental profile) gas velocity 
dispersions for all the spaxels with the  H\,{\sc ii}-region-like spectra  $\sigma_{gas,HIIobs}$ in km/s. 
}
\begin{center}
\begin{tabular}{cccccccccc} \hline \hline
Name                  &
d                     &
M$_{sp}$               &
$i$                   &
R$_{INT}$              &
R$_{\sigma_{gas}}$        &
logL$_{INT,Ha}$        &           
log$\Sigma_{Ha,C}$      &           
log$\sigma_{gas,c}$     &
log$\sigma_{gas,HIIobs}$     \\
                      &
(Mpc)                 &
(M$_{\sun}$)           &
($\degr$)             &
(kpc)                 &           
(kpc)                 &
(erg/s)               &           
(L$_{\sun}/pc^{2}$)     &           
km/s                  &           
km/s             \\   \hline
 8077 12704 &   97.2 &   10.38 &  31.0 &    1.77 &    2.24 &  39.41 &  -1.200 &   1.744 &   1.844 \\
 8080 06104 &  147.4 &   11.09 &  29.6 &    1.22 &    1.61 &  40.06 &  -0.344 &   1.959 &   1.862 \\
 8085 12704 &  123.5 &   10.66 &  60.6 &    1.84 &    2.25 &  39.98 &  -0.498 &   1.915 &   1.859 \\
 8135 12703 &  210.1 &   10.69 &  55.9 &    2.43 &    3.32 &  39.72 &  -1.326 &   1.849 &   1.837 \\
 8137 12703 &  158.7 &   10.89 &  45.1 &    2.25 &    2.89 &  40.12 &  -0.618 &   1.927 &   1.853 \\
 8138 12704 &  128.3 &   11.36 &  54.3 &    1.13 &    2.34 &  40.06 &  -0.096 &   2.140 &   1.872 \\
 8141 12704 &  199.1 &   10.63 &  41.4 &    1.41 &    3.62 &  39.58 &  -0.830 &   1.734 &   1.852 \\
 8329 06103 &  135.3 &   10.57 &  54.8 &    1.48 &    2.13 &  39.95 &  -0.530 &   1.883 &   1.895 \\
 8623 06104 &  398.5 &   11.28 &  34.1 &    2.32 &    4.36 &  41.05 &   0.265 &   2.094 &   1.871 \\
 8990 09102 &  139.8 &   10.69 &  49.0 &    0.85 &    2.21 &  39.69 &  -0.272 &   1.897 &   1.870 \\
 9027 12701 &  136.3 &   11.00 &  43.7 &    1.16 &    1.82 &  40.09 &  -0.139 &   1.896 &   1.867 \\
 9039 06102 &  261.3 &   11.34 &  41.2 &    2.64 &    3.49 &  40.63 &  -0.242 &   2.159 &   1.903 \\
 9041 12702 &  137.6 &   10.88 &  59.7 &    0.84 &    2.84 &  39.52 &  -0.421 &   1.819 &   1.882 \\
 9042 12702 &  309.3 &   11.02 &  53.4 &    2.51 &    4.88 &  40.79 &   0.042 &   2.031 &   1.867 \\
 9500 09102 &  180.1 &   10.21 &  40.8 &    1.20 &    3.28 &  39.71 &  -0.590 &   1.630 &   1.872 \\
 9506 09102 &  113.5 &   10.29 &  67.2 &    1.00 &    3.99 &  39.41 &  -0.662 &   1.483 &   1.871 \\
 9868 12705 &  306.1 &   11.13 &  48.1 &    1.98 &    4.83 &  40.61 &  -0.069 &   2.040 &   1.884 \\
\hline
\end{tabular}\\
\end{center}
\begin{flushleft}
\end{flushleft}
\end{table*}

\section{Circumnuclear regions of different BPT types}

\subsection{BPT types}

The intensity of strong, easily measured lines can be used to separate  
different types of emission-line objects according to their main  
excitation mechanism (i.e. starburst or AGN). \citet{Baldwin1981}  proposed a diagram 
(BPT classification diagram) where 
the excitation properties of H\,{\sc ii} regions are studied by 
plotting the low-excitation [N\,{\sc ii}]$\lambda$6584/H$\alpha$ (= X$_{BPT}$) 
line ratio against the high-excitation [O\,{\sc iii}]$\lambda$5007/H$\beta$ (= Y$_{BPT}$)  line ratio. 

\citet{Kauffmann2003}  found an empirical demarcation line between the star forming 
and the AGN spectra in the BPT diagram (solid line in panel c of Fig.~\ref{figure:bpt-definition}).
This demarcation line  can be interpreted as
the upper limit of pure star forming  spectra \citep{Davies2014a,Davies2014b,Davies2016}. 
The spectra located left (below) the demarcation line of \citet{Kauffmann2003} will be
referred to as the SF-like or  H\,{\sc ii} region-like spectra (blue points in panel c of Fig.~\ref{figure:bpt-definition}).
\citet{Kewley2001}  found a theoretical demarcation line between the star forming
  and the AGN spectra in the BPT diagram (short-dashed line in panel c of Fig.~\ref{figure:bpt-definition}).
This demarcation line can be interpreted as the lower limit of pure AGN spectra.  
The spectra located right (above) the demarcation line of \citet{Kewley2001} will be referred to 
as the AGN-like spectra  (black points in panel c of Fig.~\ref{figure:bpt-definition}).
The spectra located between the demarcation lines of \citet{Kauffmann2003} and \citet{Kewley2001}
will be referred to as the intermediate (INT) spectra (red points in panel c of Fig.~\ref{figure:bpt-definition}). 
In the literature, those spectra are also referred  to as composite or 
transition spectra \citep{Davies2014a,Pons2014,Pons2016}.
The long-dashed line in panel c of Fig.~\ref{figure:bpt-definition} is the dividing line between
Seyfert galaxies and LINERs defined by \citet{CidFernandes2010}. 

Here we examine the circumnuclear regions in MaNGA galaxies. Using the BPT classification diagram,
the circumnuclear regions are divided into three BPT types: the AGN type, the SF (or  H\,{\sc ii})
type, and the intermediate (INT) type.

\subsection{Circumnuclear regions of the AGN type}

\subsubsection{Characteristics of the circumnuclear region}

\begin{figure}
\begin{center}
\resizebox{1.00\hsize}{!}{\includegraphics[angle=000]{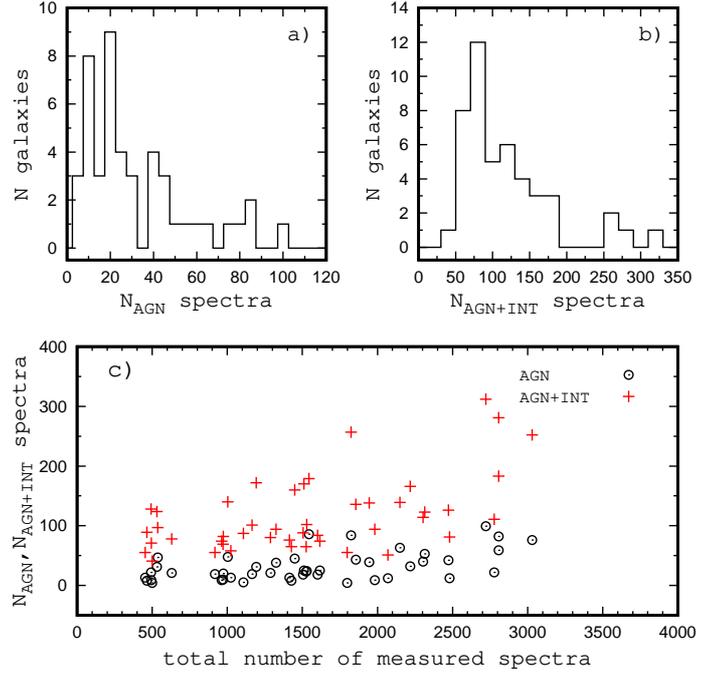}}
\caption{   
  Histograms of the numbers of the AGN-like spectra N$_{AGN}$ ({\em panel} $a$)
  and AGN-like+intermediate spectra N$_{AGN+INT}$ ({\em panel} $b$)  in the circumnuclear regions
  in galaxies with the circumnuclear regions of the AGN types. 
{\em Panel} $c$:  N$_{AGN}$ (circles) and  N$_{AGN+INT}$ (plus signs) 
as a function of total number of spectra measured in the galaxy. 
}
\label{figure:n-n}
\end{center}
\end{figure}

Panel a of Fig.~\ref{figure:bpt-definition}  shows the location of the spaxels with the AGN-like,
H\,{\sc ii}-region-like, and intermediate spectra on the image of the galaxy M-7495-12704, where the
circumnuclear region is of the AGN type. The BPT classification of each spectra is color-coded.
The spaxels with the AGN-like (intermediate) spectra in the circumnuclear region (center) and
in the disc are shown by different colors. The circle marks the kinematic center of the galaxy.
Panel b  of Fig.~\ref{figure:bpt-definition} shows a map of the observed (non corrected for instrumental profile)
gas velocity dispersion. The value of gas velocity dispersion is color-coded.  
Fig.~\ref{figure:rg-param-example}  shows the radial variation of several measured properties of this galaxy. 

The circumnuclear region with an AGN-induced radiation can be specified by a
number of parameters as indicated below.
The luminosity of the circumnuclear region with an AGN-induced radiation will  be specified by the
following parameters.  The $L_{AGN,{\rm H}\alpha}$ luminosity is defined as the sum of the H$\alpha$ luminosities
of the spaxels with AGN-like spectra in the circumnuclear region. 
The $L_{INT,{\rm H}\alpha}$ luminosity is defined as the sum of the H$\alpha$ luminosity
of the spaxels with AGN-like and intermediate spectra in the circumnuclear region.
The $L_{AGN,OIII}$ and  $L_{INT,OIII}$ luminosities in the [O\,{\sc iii}]$\lambda\lambda$4959,5007 
lines are determined in a similar way using the [O\,{\sc iii}]$\lambda$5007 line measurement and 
adopting $L_{[OIII]\lambda\lambda4959,5007}$ = 1.3$L_{[OIII]\lambda5007}$ \citep{Storey2000}.
It is believed that the H$\alpha$ emission within the circumnuclear region is likely
to be strongly contaminated by star formation, whereas the [O\,{\sc iii}] emission is less
contaminated and is therefore a better indicator of the AGN luminosity.

The size of the circumnuclear region with an AGN-induced radiation will be specified by 
two radii. Since the shape of the circumnuclear region with an AGN-induced radiation
can deviate significantly from the ring on the deprojected plane (ellipses in the observed
plane) then we define the radius $R_{AGN}$ as the radius of the ring with an area equal
to the area covered by the spaxels with the AGN-like spectra. The radius $R_{AGN}$
specifies the size of the zone where the AGN-induced radiation makes a dominant contribution to
the emission spectra.
Similarly, we define the radius $R_{INT}$ as the radius of the ring with the area equal to the
area covered by the spaxels with the AGN-like and intermediate spectra.
The radius $R_{INT}$ specifies the size of the zone where the AGN-induced radiation makes a contribution to
the emission spectra large enough such that the spectra are classified as  INT type.
In other words, the radius $R_{INT}$ is the radius of influence of AGN. 

The determination of the radius $R_{INT}$ can meet the following difficulty.
It was noted above that the discs of many galaxies involve low-ionization emission-line regions (LIER)
excited by the HOLMES. If the LIER adjoins to the LINER then the central area covered by the spaxels with
the AGN-like and the intermediate spectra (and, consequently, the radius $R_{INT}$) will be overestimated.
This seems take place in the galaxy M-7495-12704, the north part of the circumnuclear region
may be an adjoining LIER (panel a of Fig.~\ref{figure:bpt-definition}).

Panel c of Fig.~\ref{figure:rg-param-example}  shows that the observed
(non corrected for the instrumental profile)  gas velocity dispersion 
decreases with galactocentric distance up to some radius $R_{\sigma_{gas}}$
and remains approximately constant beyond this radius. 
The radius of the region with enhanced gas velocity dispersion $R_{\sigma_{gas}}$ 
is estimated for each galaxy. The radial distribution of the  $\sigma_{gas,obs}$ is 
fitted with a broken line, and the radius where the break occurs is adopted as the $R_{\sigma_{gas}}$. 
To minimize the influence of errors in the  H$\alpha$ line width measurements
in the spectra of individual spaxels on the determination of the  $R_{\sigma_{gas}}$,
the median values of the  $\sigma_{gas,obs}$ within rings with width of 1 spaxel
are determined (panel a of Fig.~\ref{figure:r}),  and the broken line fit to those
median values is obtained.

The gas velocity dispersion within the circumnuclear region can be also specified by the 
maximum value of the gas velocity dispersions.
Again, to minimize the influence of errors in the  H$\alpha$ line width measurements
in the spectra of individual spaxels on this parameter, we specify the gas velocity dispersion
within the circumnuclear region  by the mean value of the gas velocity dispersions in
five spaxels in the circumnuclear region with higher $\sigma_{gas}$ and will refer to this parameter as 
the central gas velocity dispersion $\sigma_{gas,c}$.
The median value of the observed (non corrected for the instrumental profile) gas
velocity dispersion for all the spaxels of the H\,{\sc ii}-region-like spectra,
$\sigma_{median,HIIobs}$, is also obtained for each galaxy. 

The radial distribution of the  observed stellar velocity dispersion $\sigma_{star,obs}$ 
is also fitted by the broken line, and the break radius is adopted as the $R_{\sigma_{star}}$ 
(panel b of Fig.~\ref{figure:r}). 
The uncertainty in  $\sigma_{star}$ is much larger than that in   $\sigma_{gas}$  (panels a and b 
of Fig.~\ref{figure:r}).   There are no a plateaus in the  $\sigma_{star}$ -- R diagrams, and t
he observed stellar velocity dispersions are below the MaNGA instrumental 
dispersion in many cases \citep{Westfall2019}. 
A reliable estimation of the  $R_{\sigma_{star}}$ value was not obtained 
for a several galaxies in our sample. 

The measured flux in the  H$\alpha$ line per spaxel $F_{{\rm H}\alpha}$ in units 10$^{-17}$ erg/s/cm$^2$/spaxel
is converted into  surface brightness in the H$\alpha$ line $\Sigma_{{\rm H}\alpha}$.  
The value of the surface brightness is corrected for the galaxy inclination. 
The central surface brightness  $\Sigma_{{\rm H}\alpha,c}$  is specified by the
mean values of $\Sigma_{{\rm H}\alpha}$  for the same five spaxels used for estimation of the $\sigma_{gas}$. 
The central surface brightness  in the [O\,{\sc iii}]$\lambda\lambda$4959,5007 
lines is also determined adopting again  $F_{[OIII]\lambda\lambda4959,5007}$ = 1.3$F_{[OIII]\lambda5007}$.  

Thus, the luminosities $L_{AGN}$ and $L_{INT}$ in the H$\alpha$ and [O\,{\sc iii}]$\lambda\lambda$4959,5007
emission lines, the radii  $R_{AGN}$,  $R_{INT}$,   $R_{\sigma_{gas}}$, and  $R_{\sigma_{star}}$,  
the central gas velocity dispersion $\sigma_{gas,c}$,
the central  surface brightnesses in the H$\alpha$ line $\Sigma_{{\rm H}\alpha,c}$
and in the [O\,{\sc iii}]$\lambda\lambda$4959,5007 lines  $\Sigma_{OIII,C}$, and 
the median value of the observed (non corrected for the instrumental profile) gas
velocity dispersions in all the spaxels for the H\,{\sc ii}-region-like spectra
$\sigma_{median,HIIobs}$ were determined for the circumnuclear regions of the AGN type 
for our sample of MaNGA galaxies.
The numbers of the AGN-like and intermediate spectra within the circumnuclear regions
and the total numbers of measured spectra in the galaxies  with the circumnuclear regions of
the AGN type are presented in Fig.~\ref{figure:n-n}.

\begin{table*}
  \caption[]{\label{table:cc}
Coefficients for the linear relation $y = ax + b$ between parameters of the circumnuclear
regions of the AGN type. The columns give:
the definitions of $y$ and $x$ (columns 1 and 2),
the coefficients $a$ and $b$ (columns 3 and 4),
the correlation coefficient (column 5),
and the mean value of the scatter in $y$ (column 6).
The respective diagrams are shown in  Fig.~\ref{figure:svc-l-agn}  and  Fig.~\ref{figure:r-l-agn}. 
}
\begin{center}
\begin{tabular}{llcccc} \hline \hline
$y$                  &
$x$                  &
$a$                  &
$b$                  &
correlation          &
scatter             \\  
                     &
                     &
                     &
                     &
coefficient          &
in $y$              \\  \hline 
\multicolumn{6}{l}{Relations in  Fig.~\ref{figure:svc-l-agn}  }    \\           
log$L_{AGN,{\rm H}\alpha}$        &  $\sigma_{gas,c}$                &  0.0095$\pm$0.0021        &  38.68$\pm$0.23       &  0.555 &  0.332    \\
log$L_{INT,{\rm H}\alpha}$        &  $\sigma_{gas,c}$                &  0.0050$\pm$0.0016        &  39.79$\pm$0.17       &  0.424 &  0.249    \\
log$L_{AGN,OIII}$               &  $\sigma_{gas,c}$                &  0.0100$\pm$0.0023        &  38.46$\pm$0.25       &  0.539 &  0.364    \\
log$L_{INT,OIII}$               &  $\sigma_{gas,c}$                &  0.0060$\pm$0.0018        &  39.30$\pm$0.19       &  0.454 &  0.276    \\
\multicolumn{6}{l}{Relations in Fig.~\ref{figure:r-l-agn} }    \\           
log$R_{AGN}$                   &  log$L_{AGN,{\rm H}\alpha}$       &  0.324$\pm$0.060          & -12.68$\pm$2.40       &  0.629 &  0.160    \\
log$R_{AGN}$                   &  log$L_{AGN,OIII}$               &  0.334$\pm$0.051         & -13.03$\pm$2.01       &  0.703 &  0.146    \\
log$R_{INT}$                   &  log$L_{INT,{\rm H}\alpha}$       &  0.302$\pm$0.074          & -11.66$\pm$3.00       &  0.522 &  0.136    \\
log$R_{INT}$                   &  log$L_{INT,OIII}$               &  0.288$\pm$0.064         & -11.01$\pm$2.55       &  0.562 &  0.131    \\
log$R_{AGN}$                   &  log$\sigma_{gas,c}$            &  1.102$\pm$0.247         &  -2.033$\pm$0.496      &  0.558 &  0.171    \\
log$R_{INT}$                   &  log$\sigma_{gas,c}$            &  0.628$\pm$0.210         &  -0.758$\pm$0.421      &  0.412 &  0.145    \\
\hline
\end{tabular}
\end{center}
\end{table*}

\subsubsection{AGN-SF mixing}

The positions of the AGN-like and intermediate spectra of the spaxels in 
the circumnuclear region occupy a rather narrow band in the BPT diagram, see
panel c in Fig.~\ref{figure:bpt-definition}. This sequence on the BPT diagram 
can be fitted by the linear relation
\begin{equation}
Y_{BPT} = a\; X_{BPT} + b .
\label{equation:xybpt}
\end{equation}
Such relation (coefficients $a$ and $b$ in Eq.~\ref{equation:xybpt}) 
was obtained for all galaxies with the circumnuclear regions of the AGN type. 
It should be emphasized that the spaxels with the intermediate spectra
in the disc (beyond the circumnuclear region) do not follow this relation, see
panel d in Fig.~\ref{figure:bpt-definition}. 

\citet{KauffmannHeckman2009} created a set of empirical mixing line trajectories in the BPT 
diagram by averaging the emission-line fluxes of  star-forming galaxies and  AGNs 
in different proportions. The decrease of the fraction of the luminosity attributed to SF 
results in a shift from the locus of pure star-forming galaxies in the BPT diagram. 
The greater the fraction of the line emission excited by AGN activity, the further 
along the mixing trajectory a spectrum will lie. 

\citet{Davies2014a,Davies2014b,Davies2016}  found that spectra of individual spaxels 
of some AGN host galaxies  form a tight sequence on the BPT diagram, which is similar 
to the mixing line trajectories constructed by \citet{KauffmannHeckman2009} for ``global''
spectra. The diagnostic line ratios in the spaxel spectra change smoothly with radius, 
from AGN-like  ratios in the spaxel spectra at the galactic center, to 
H\,{\sc ii}-like  ratios in the spaxel spectra at larger galactocentric distances. 
\citet{Davies2014a,Davies2014b,Davies2016} concluded that this provides strong evidence 
in favor of variations of diagnostic line ratios  with radius being primarily driven by 
variations in the fraction of the line emission excited by AGN activity. 

Panels c and d in Fig.~\ref{figure:rg-param-example} show the variations in the
diagnostic line ratios [N\,{\sc ii}]$\lambda$6584/H$\alpha$ and 
[O\,{\sc iii}]$\lambda$5007/H$\beta$ in the spectra of spaxels from 
the circumnuclear region of the galaxy M-7495-12704 as a function of 
galactocentric distance, respectively. Inspection of panel b in Fig.~\ref{figure:bpt-definition} 
together with panels c and d in Fig.~\ref{figure:rg-param-example}, show clearly 
that the diagnostic line ratios of the spaxel spectra change smoothly with radius, 
this is, the spaxel spectra of the circumnuclear region lie along the mixing line trajectories
of the BPT diagram, where the fraction of the line emission excited by AGN activity decreases with radius.  
It should be emphasized that the same radial behavior in the diagnostic lines was obtained 
for every one of the galaxies in our sample where the circumnuclear region is of the AGN-type. 

The diagnostic line ratios in the spaxel spectra change smoothly with the radius within
the circumnuclear region, and the gas velocity dispersion decreases with galactocentric
distance  within the circumnuclear region, that is, the diagnostic line ratios correlate with 
gas velocity dispersion. It is worth noting that such a correlation is not only related to
AGN excited regions. A correlation between gas velocity dispersion and emission-line
ratio is revealed in the ultraluminous and luminous infrared galaxies in the absence of any
contribution from an AGN. This correlation is interpreted as a signature of shock excitation
where shocks are driven by galaxy mergers or interactions \citep{Monreal2006,Monreal2010,Rich2014,Rich2015}.

The relations that were obtained above between $X_{BPT}$ and $Y_{BPT}$, Eq.~(\ref{equation:xybpt}), 
describe the mixing line trajectories on the BPT diagram for galaxies with the circumnuclear 
region of the AGN-type, see panel c in Fig.~\ref{figure:bpt-definition}. 
Next, we examine the dependences between the slope of the mixing line trajectory  
on the BPT diagram with other properties of the galaxy
(the slope of the relation 
$Y_{BPT}$ = $a \times X_{BPT}$ + $b$ as a function   of the stellar mass of the galaxy,
H$\alpha$ luminosity of the AGN zone,  central gas velocity dispersion,
and central surface brightness in the H$\alpha$ line). 
We find that there is not a significant corrrelation between the slope of the  mixing line
trajectory on the BPT diagram with other parameters of the galaxy.

\subsubsection{Gas velocity dispersion as an indicator of  AGN activity}

\begin{figure}
\begin{center}
\resizebox{1.00\hsize}{!}{\includegraphics[angle=000]{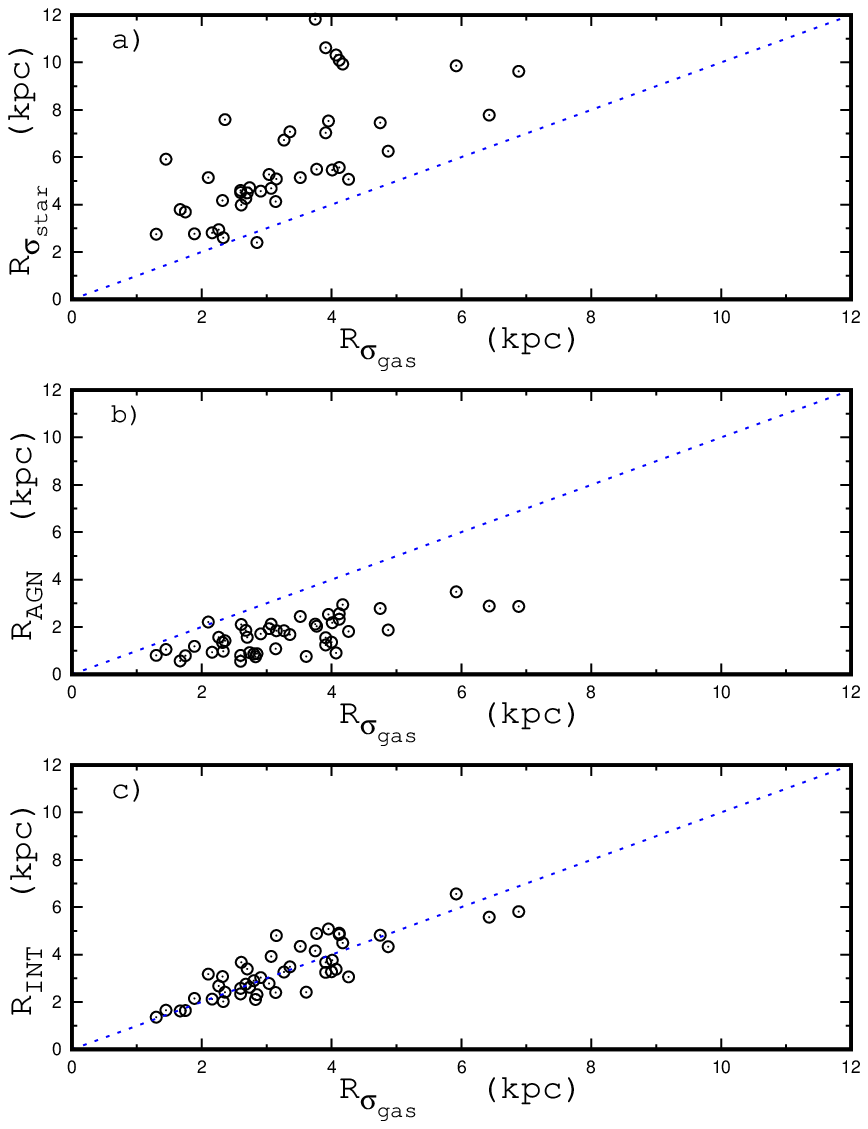}}
\caption{
Comparison between the radii of different zones for MaNGA galaxies with an
 AGN-type circumnuclear regions. 
The panels show 
the radius of the zone of the enhanced stellar velocity dispersion $R_{\sigma_{star}}$  ({\em panel} $a$),
the radius of the zone where the dominant contribution is  AGN-type radiation $R_{AGN}$  ({\em panel} $b$),
and the radius of the zone with AGN influence on the radiation $R_{INT}$  ({\em panel} $c$)  
as a function of the radius with an enhanced gas velocity dispersion $R_{\sigma_{gas}}$.
In each panel, the circles denote  individual galaxies. 
The dashed line represents the one-to-one relation. 
}
\label{figure:rrr-agn}
\end{center}
\end{figure}

\begin{figure}
\begin{center}
\resizebox{1.00\hsize}{!}{\includegraphics[angle=000]{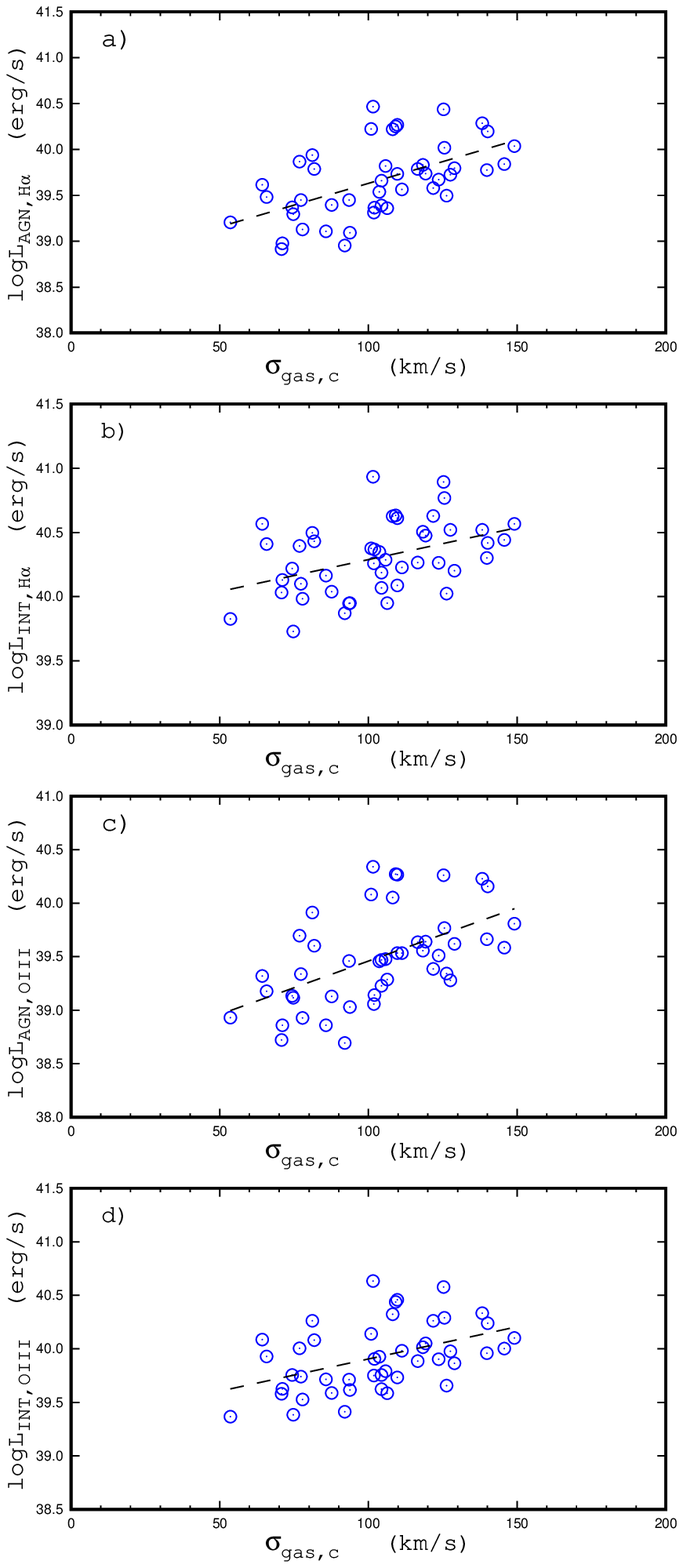}}
\caption{
Relations between the luminosity of the circumnuclear region of the AGN type 
and central gas velocity dispersion. 
{\em Panel} $a$ shows the H$\alpha$ line luminosity for the region where the dominant
contribution to the radiation is an AGN, $L_{AGN,{\rm H}\alpha}$, 
as a function of the central gas velocity dispersion $\sigma_{gas,c}$. 
{\em Panel} $b$ shows the H$\alpha$ line luminosity of the zone of influence 
of the AGN, $L_{INT,{\rm H}\alpha}$,  as a function of central gas velocity dispersion. 
{\em Panels} $c$ and $d$ show the same as {\em panels} $a$ and $b$
but for the luminosity in the [O\,{\sc iii}]$\lambda\lambda$4959,5007 emission lines.
}
\label{figure:svc-l-agn}
\end{center}
\end{figure}

The characteristics of a galaxy with an AGN can be divided into two groups.
The parameters of the first group (e.g., stellar mass, $M_{sp}$) specify the host galaxy. 
The parameters of the second group (e.g., luminosity, $L_{INT}$, or radius of the
influence of the AGN, $R_{INT}$)  specify the level of the AGN activity.

Faber-Jackson relations link the luminosity (mass) and the stellar velocity dispersion in  elliptical
galaxies and in the bulges of disc galaxies \citep{Faber1976,Kormendy1985}.  
It is also established  that the mass of supermassive black holes correlates tightly with the 
stellar velocity dispersion of the host galaxy 
\citep[e.g.,][]{Ferrarese2000,Tremaine2002,King2015,Sahu2019}. 
Therefore, the stellar velocity dispersion specifies the host galaxy and
belongs to the first group of parameters.

It was suggested \citep{Nelson1996} to use the width of the [O\,{\sc iii}]$\lambda$5007 
emission line as a surrogate for $\sigma_{star}$ for Seyfert galaxies. 
\citet{Bennert2018} carried out a direct comparison between $\sigma_{star}$ and $\sigma_{OIII}$ 
and found that there is no correlation  between $\sigma_{star}$ and $\sigma_{OIII}$. 
On the other hand, \citet{Ilha2019} have found that the fractional difference
between gas and star velocity dispersions
$\sigma_{frac}$ = ($\sigma_{gas}$ -- $\sigma_{star}$)/$\sigma_{star}$ 
can be used as an indicator of AGN activity.
Here we examine the relationships between the gas velocity dispersion  and 
other parameters for our sample of  MaNGA galaxies. 
Among other things, this can tell us if the gas velocity dispersion 
in the LINERs is an indicator of the (gaseous) bulge (belongs to the first group
of parameters) or the AGN activity (belongs to the second group of parameters).

\begin{figure*}
\begin{center}
\resizebox{1.00\hsize}{!}{\includegraphics[angle=000]{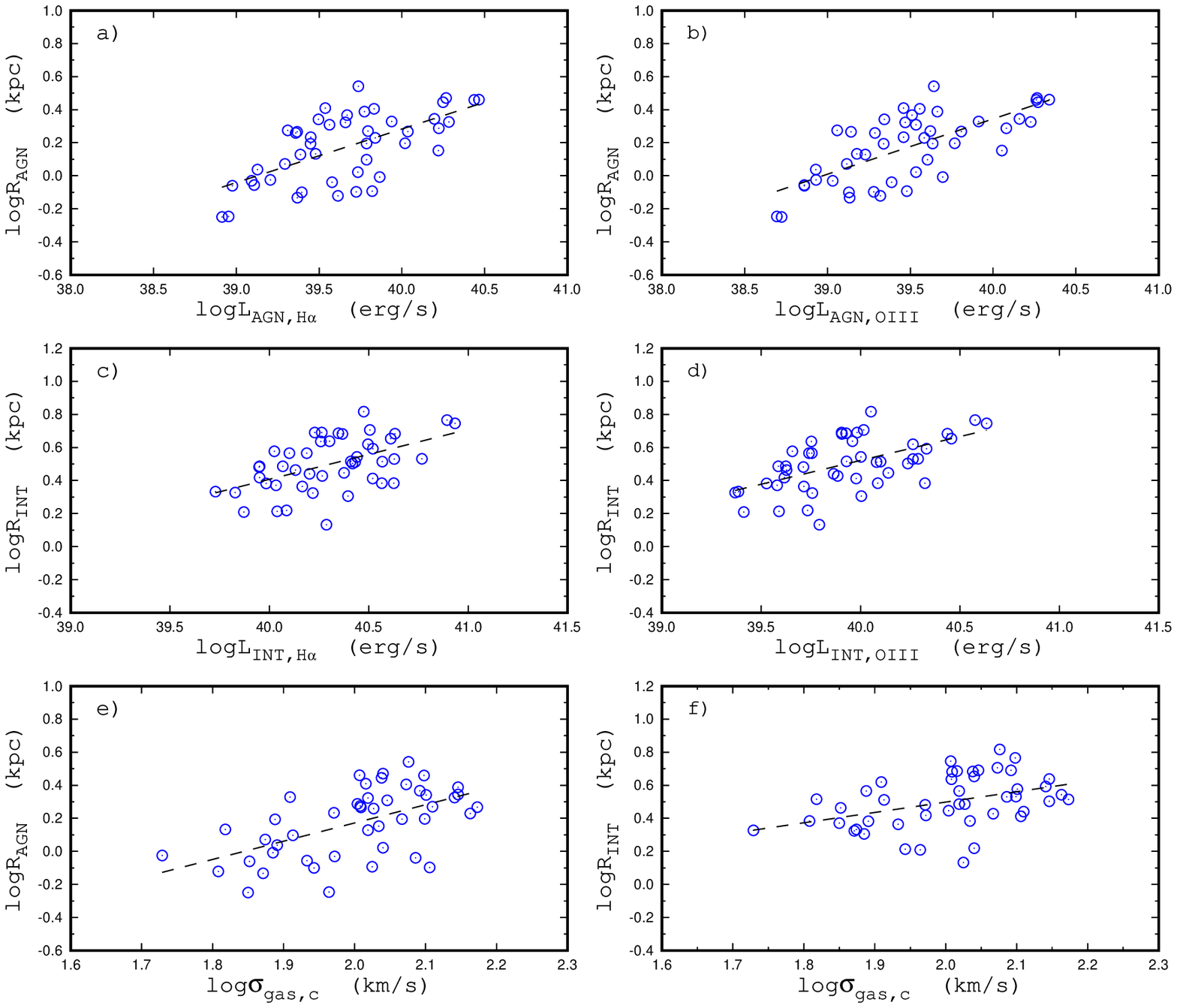}}
\caption{
Relations between the radius and other parameters of the circumnuclear 
region of the AGN-type. 
Panels a, b, and e show the radius of the zone of dominant contribution 
of the AGN to the radiation, $R_{AGN}$, as a function of the luminosity of 
this zone in the H$\alpha$ emission line (panel a), in the 
[O\,{\sc iii}]$\lambda\lambda$4959,5007 emission lines (panel b), and as a function of 
the central gas velocity dispersion (panel e). 
Panel c, d, and f show the same as panels a, b, e (respectively) but for the radius of the zone of 
the AGN influence on the radiation, $R_{INT}$. The points in each panel denote  individual
galaxies. The line is the best linear  fit to those data (see Table \ref{table:cc}).
}
\label{figure:r-l-agn}
\end{center}
\end{figure*}

The energy generation by an AGN (and consequently its influence on the circumnuclear region)
depends on mass of the black hole and acretion rate. A significant  correlations
exist between the mass of the black hole and some properties of the host galaxy,
such as stellar mass or stellar velocity dispersion. Therefore, examination of correlations
between $\sigma_{gas}$ and those properties cannot provide the choice whether
the $\sigma_{gas}$ is a tracer of AGN activity (related to the black hole mass) or is a tracer
of gaseous bulge (related to the host galaxy properties). At the same time, the existence of
correlations between  $\sigma_{gas}$ and direct characteristics of the AGN activity,
such as AGN luminosity and radius of AGN influence, can be considered as a significant evidence
in favor of that the $\sigma_{gas}$ is a tracer of AGN activity.

Fig.~\ref{figure:rrr-agn} shows the comparison between the radius of the zone 
with enhanced gas velocity dispersion $R_{\sigma_{gas}}$ and the radii of other 
zones in the galaxies. Inspection of panel a in Fig.~\ref{figure:rrr-agn} shows 
that the radius of the zone with enhanced gas velocity dispersion $R_{\sigma_{gas}}$ 
is smaller than the radius of the zone with enhanced stellar velocity dispersion 
$R_{\sigma_{star}}$.  Examination of panel b in Fig.~\ref{figure:rrr-agn} shows that 
the $R_{\sigma_{gas}}$ exceeds the size of the zone with dominant contribution of AGN 
to the radiation $R_{AGN}$. Panel c of Fig.~\ref{figure:rrr-agn} shows  
that  $R_{\sigma_{gas}}$ is very similar to the radius of the zone of the influenc of the 
AGN on the radiation $R_{INT}$. 

Thus, the radius of the zone with enhanced gas velocity dispersion 
$R_{\sigma_{gas}}$ is related to the radius of the zone of the influence of the 
AGN on the radiation $R_{INT}$. 

Fig.~\ref{figure:svc-l-agn} shows the luminosity of the H$\alpha$  and
the [O\,{\sc iii}]$\lambda\lambda$4959,5007 emission lines for the zone with the
dominant AGN contribution to the radiation, $L_{AGN}$, and of the zone  
of the influence of AGN, $L_{INT}$,  as
a function of the central gas velocity dispersion $\sigma_{gas,c}$.
In each panel, the circles correspond to individual galaxies,
and the line is the best fit to those data.
Coefficients for the linear correlations of the type  log$L$ = $a\sigma_{gas,c} + b$,
the correlation coefficient, and the mean value of the scatter in  log$L$ for each diagram
are listed in Table \ref{table:cc}.
Every one of those luminosities  ($L_{AGN,{\rm H}\alpha}$, $L_{AGN,OIII}$, 
$L_{INT,{\rm H}\alpha}$, $L_{INT,OIII}$) is an indicator of  AGN activity. 
Examination of Fig.~\ref{figure:svc-l-agn} shows that there is a 
correlation between $\sigma_{gas,c}$ and luminosity of each kind. 
This provides  evidence in favor of that the central gas velocity dispersion 
$\sigma_{gas,c}$  can serve as an indicator of  AGN activity. 

It is known that the radius of the zone ionized by the AGN correlates 
with its luminosity in  the [O\,{\sc iii}]$\lambda$5007 emission line
for the Seyfert galaxies \citep[][among many others]{Schmitt2003,Chen2019,doNascimento2019}.
In Fig.~\ref{figure:r-l-agn} we show the radius  of the zone where an AGN makes the 
dominant contribution to the radiation, $R_{AGN}$, as a function 
of  the H$\alpha$ luminosity for the same zone, $L_{AGN,{\rm H}\alpha}$, (panel a),
and as a function of  the [O\,{\sc iii}]$\lambda\lambda$4959,5007 
luminosity (panel b) for our sample of LINERs. 
The radius  of the zone of the influence of the AGN on the radiation, $R_{INT}$, 
is plotted as a function of the H$\alpha$ luminosity, for the same zone, $L_{INT,{\rm H}\alpha}$ 
in panel c of Fig.~\ref{figure:r-l-agn}, and as a function of  the [O\,{\sc iii}]$\lambda\lambda$4959,5007 luminosity  
in panel d of Fig.~\ref{figure:r-l-agn}.  
One can see the correlation between $R$ and $L$ in each diagram.
The coefficients in the linear correlations  $y =ax + b$, the correlation coefficient,
and the mean value of the scatter in $y$ for each diagram are listed in Table \ref{table:cc}.

Panel e in Fig.~\ref{figure:r-l-agn} shows the radius $R_{AGN}$ as a function of the
central gas velocity dispersion, while panel f shows the  $R_{INT}$ -- $\sigma_{gas,c}$ diagram.
Both, the $R_{AGN}$ and the $R_{INT}$ correlate with  $\sigma_{gas,c}$. 

Thus, the consideration of our sample of galaxies with the circumnuclear regions 
of the AGN type (LINERs) results in the following conclusions. \\
The spectra of individual spaxels in the circumnuclear region occupy  
a narrow band on the BPT diagram, this is, they lie along the AGN-SF mixing line
trajectory. The diagnostic line ratios in the spaxels spectra within the radius $R_{INT}$ 
change smoothly with radius, from AGN-like line ratios in the spaxels spectra at 
the galactic center, to H\,{\sc ii}-region-like line ratios in the spaxels spectra at 
galactocentric distances  larger than $R_{INT}$. 
This is in agreement with the widely used paradigm that the LINERs are excited by the AGNs. \\
The radius of the AGN influence can be specified by the radius 
$R_{INT}$, and the strength of the AGN activity can be specified by the luminosity 
of this region in the H$\alpha$ (or [O\,{\sc iii}]$\lambda\lambda$4959,5007) 
emission lines. \\
Alternatively, the radius of the AGN influence can be specified by the radius of the zone of enhanced
gas velocity dispersion $R_{\sigma_{gas}}$ which coincides with (or at least is close to) the
radius $R_{INT}$,  and the strength of the AGN activity can be specified by the value of the
central gas velocity dispersion $\sigma_{gas,c}$.

\subsection{Circumnuclear regions of the INT type}

\begin{figure}
\begin{center}
\resizebox{0.82\hsize}{!}{\includegraphics[angle=000]{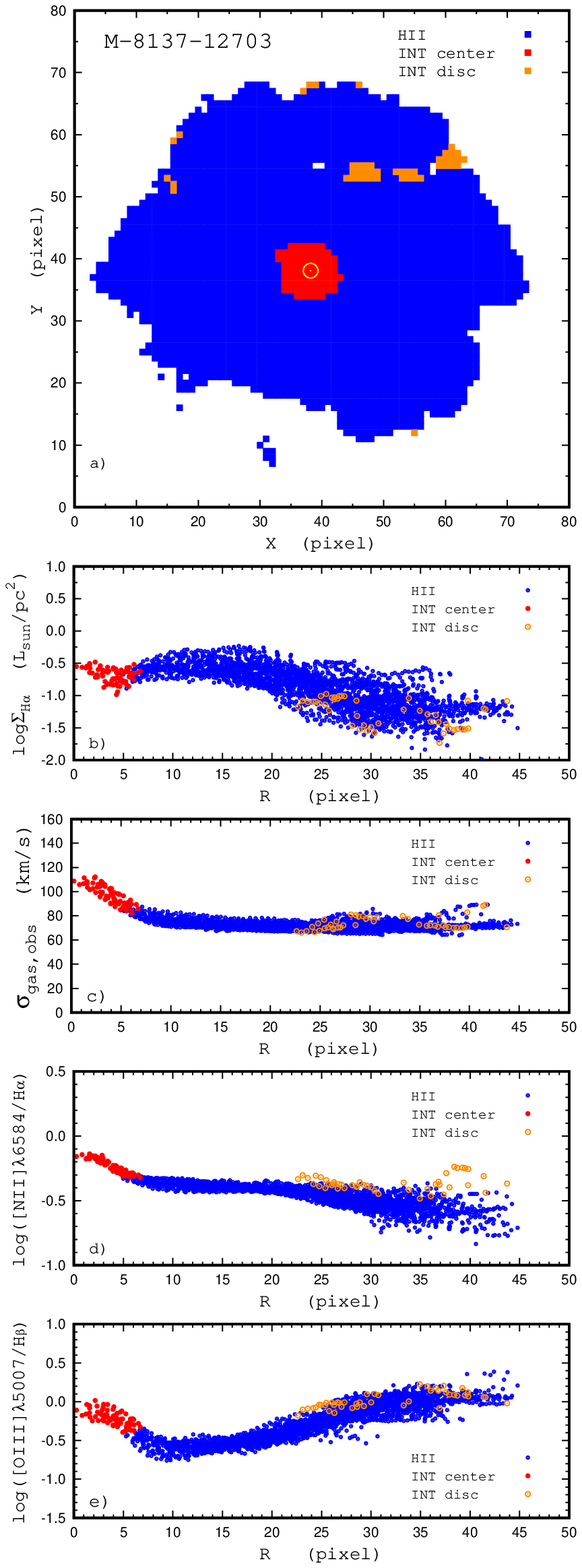}}
\caption{
Example of galaxies with the circumnuclear region classified as INT type.
{\em Panel} $a$ shows the locations of the spaxels with the H\,{\sc ii}-region-like 
and intermediate spectra on the image of  the MaNGA galaxy M-8137-12703. 
The BPT types of the spaxels spectra  are color-coded.
The spaxels with the intermediate spectra in the circumnuclear region (center) and
in the disc are shown by different colors.
The circles marks the kinematic center of the galaxy.
{\em Panels} $b$ -- $e$ show the variations with radius of 
the surface brightness in the H$\alpha$ line ({\em panel} $b$),
the observed (non corrected for the instrumental profile) gas velocity dispersion ({\em panel} $c$), 
the $X_{BPT}$ = log([N\,{\sc ii}]$\lambda$6584/H$\alpha$) ({\em panel} $d$), 
and the $Y_{BPT}$ = log([O\,{\sc iii}]$\lambda$5007/H$\beta$) ({\em panel} $e$). 
In each panel, 1 pixel corresponds to 0.385 kpc.
}
\label{figure:int-example}
\end{center}
\end{figure}

\begin{figure}
\begin{center}
\resizebox{1.00\hsize}{!}{\includegraphics[angle=000]{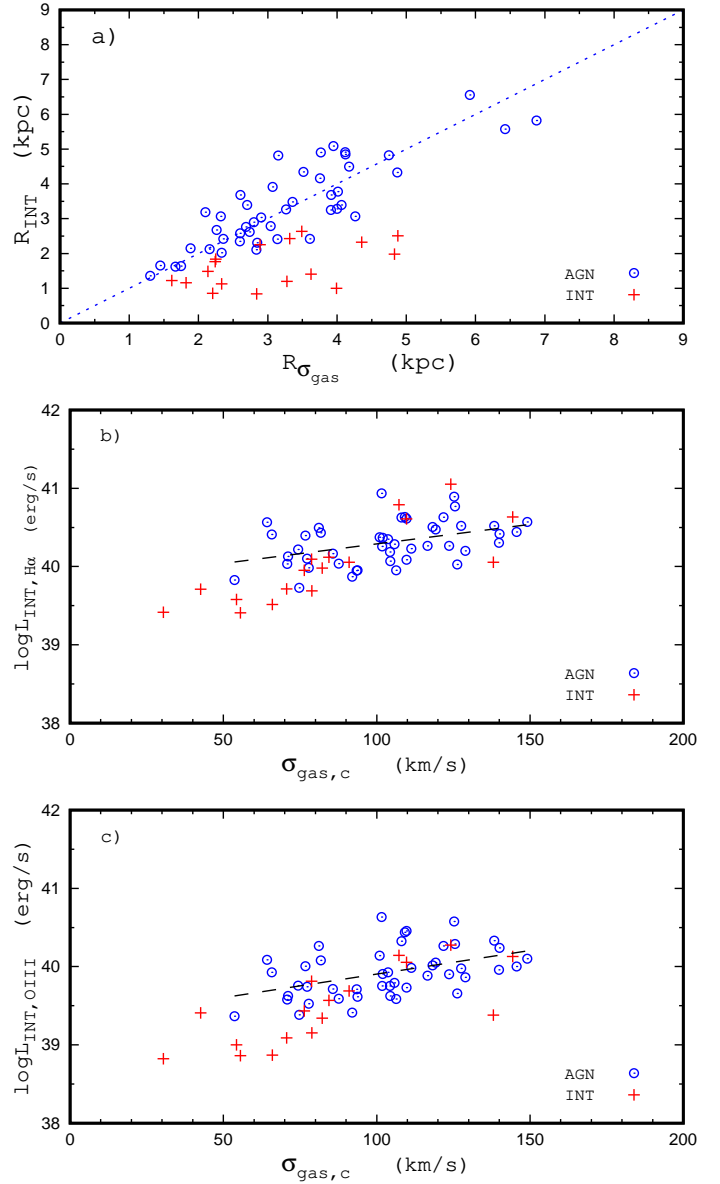}}
\caption{
Comparison of properties of the circumnuclear regions of the AGN
(blue symbols) and INT (red symbols) types.
{\em Panel} $a$ shows the $R_{INT}$ -- $R_{\sigma_{gas}}$ diagram.
The line is the perfect agreement.
{\em Panel} $b$ shows the $L_{INT,{rm H}\alpha}$ -- $\sigma_{gas,c}$ diagram.
The line is the $L_{INT,{rm H}\alpha}$ -- $\sigma_{gas,c}$ relation for  galaxies with
the circumnuclear region of the AGN type. 
{\em Panel} $c$ shows the $L_{INT,OIII}$ -- $\sigma_{gas,c}$ diagram.
The line is the $L_{INT,OIII}$ -- $\sigma_{gas,c}$ relation for  galaxies with
the circumnuclear region of the AGN type. 
}
\label{figure:r-r-int}
\end{center}
\end{figure}

\begin{figure}
\begin{center}
\resizebox{0.80\hsize}{!}{\includegraphics[angle=000]{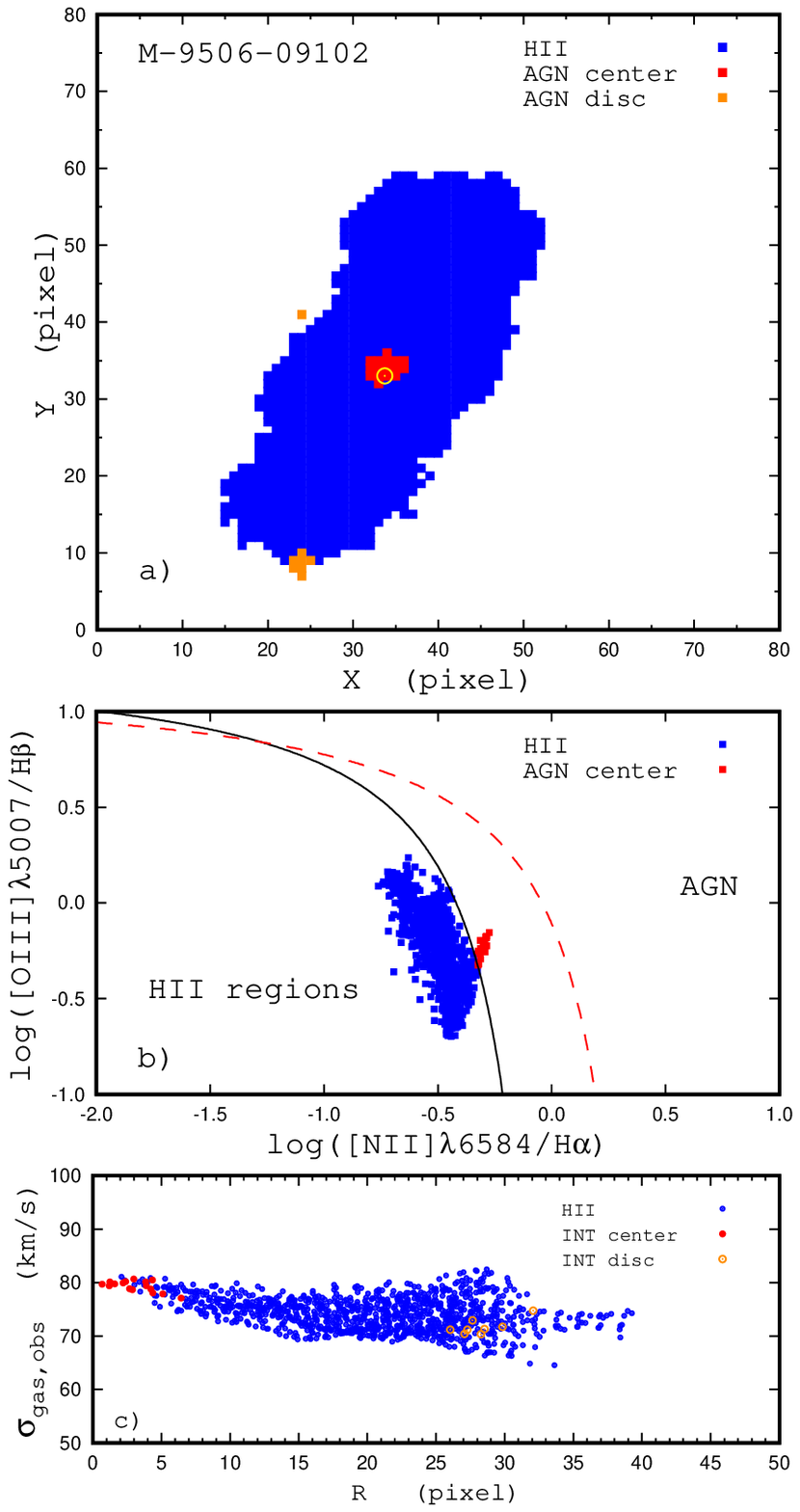}}
\caption{
Example of a galaxy with a weak circumnuclear region of the INT type. 
{\em Panel} $a$ shows the location of  spaxels with different
BPT classifications for the MaNGA galaxy M-9506-09102. 
The spaxels with  H\,{\sc ii}-region-like spectra are denoted by  blue symbols,
the spaxels with  intermediate spectra in the circumnuclear region are marked
by  red symbols, and those in the disc by  orange symbols.
The circle is the kinematic center of the galaxy. 
{\em Panel} $b$ shows the BPT diagram for the spectra of the individual spaxels.
The colors correspond to the same classification as in panel $a$.
Solid and short-dashed curves mark the demarcation line between AGNs and H\,{\sc ii}
regions defined by \citet{Kauffmann2003} and \citet{Kewley2001}, respectively.
{\em Panel} $c$ shows the variation with radius of the observed (non corrected for the
instrumental profile) gas velocity dispersion.
In panels a and c, 1 pixel = 0.275 kpc.
}
\label{figure:weak-int}
\end{center}
\end{figure}

The circumnuclear regions in a number of MaNGA galaxies are classified as  INT type. 
As an example, the upper panel of Fig.~\ref{figure:int-example} shows the locations of the spaxels with the 
H\,{\sc ii}-region-like and and intermediate spectra on the image of the galaxy M-8137-12703. 
The spaxels of the intermediate type spectra in the circumnuclear region and
in the disc are shown by different colors. The circle marks the kinematic center of the galaxy.
The other panels of Fig.~\ref{figure:int-example} show the radial distributions of  different
characteristics (such as the H$\alpha$ surface brightness, the gas velocity 
dispersion, and the BPT emission line ratios). 

The radius of the area of the influence of the AGN on the radiation $R_{INT}$,  
the radius of the zone of the enhanced gas velocity dispersion $R_{\sigma_{gas}}$, 
the values of the  $L_{INT,{\rm H}\alpha}$ and $L_{INT,OIII}$ luminosities within the $R_{INT}$, 
the central gas velocity dispersion $\sigma_{gas,c}$, the values of the central 
H$\alpha$ surface brightness, $\Sigma_{{\rm H}\alpha,c}$, and 
in the [O\,{\sc iii}]$\lambda\lambda$4959,5007 emission lines $\Sigma_{OIII,c}$,
are determined for the circumnuclear regions of the INT type 
in the same way as for the circumnuclear regions of the AGN type.
   
Here we compare the properties of the circumnuclear regions of the INT and the AGN types.
Panel a of Fig.~\ref{figure:r-r-int} shows the $R_{\sigma_{gas}}$ -- $R_{INT}$ diagram.  
The circles denote the circumnuclear regions of the AGN type, and the plus symbols 
are the circumnuclear regions of the INT type. The line is the perfect agreement. 
Inspection of panel a of Fig.~\ref{figure:r-r-int} shows that there is no 
agreement between the $R_{\sigma_{gas}}$ and the $R_{INT}$ for the  circumnuclear regions 
of the INT type, in the sense that the radius of the AGN influence on 
 the radiation $R_{INT}$, is usually less than the radius of the zone of the enhanced gas velocity 
dispersion $R_{\sigma_{gas}}$. This can be evidence in favor of that the contribution of 
the SF radiation is higher in those circumnuclear regions. Indeed, 
the position of each spaxel on the BPT diagram moves along the  mixing line trajectories 
toward  lower $X_{BPT}$ and  $Y_{BPT}$ values due to an increase of the contribution of 
the SF radiation. As a result, the positions of the central spaxels move from the 
AGN locus to the INT locus on the BPT diagram, and the position of the outermost 
spaxels (with radii close to, slightly less than, the  $R_{\sigma_{gas}}$) move 
from the INT locus to the  H\,{\sc ii} region locus.
Thus, the differences between the circumnuclear regions of the AGN and INT types can be explained
if the contribution of the SF to the radiation is higher in circumnuclear regions of the
INT type, in comparison to the circumnuclear regions of the AGN type. 

It should be noted that the enhancement of the gas velocity dispersion in
some galaxies of the INT type (with weak circumnuclear regions) 
can be low,  Fig.~\ref{figure:weak-int}. 
Therefore the uncertainty in the determinations of the value of the $R_{\sigma_{gas}}$
for  weak circumnuclear regions of the INT type can be large.  

The BPT diagram for the circumnuclear region of the INT type resembles that of the circumnuclear
region of the AGN type, with the exception that  the positions of the spaxels of the INT type
do not reach the AGN locus in the BPT diagram. This indicates again that the contribution of 
the SF to the radiation is higher in the circumnuclear regions of the INT type 
in comparison to the circumnuclear regions of the AGN type. Since the extension of the band in the
BPT diagram occupied by the spaxels of the INT type is not large enough, then that prevents 
the accurate determination of the slope of the $Y_{BPT}$ -- $X_{BPT}$ relation. 
Moreover, even the existence of the AGN-SF mixing trajectory is not indisputable if the extension
of the band in the BPT diagram is small. 

Panel b of Fig.~\ref{figure:r-r-int} shows the $L_{INT,{\rm H}\alpha}$ -- $\sigma_{gas,c}$ 
diagram. The circles denote the circumnuclear regions of the AGN-type, and the plus symbols 
denote the circumnuclear regions of the INT-type. The line is the  $L_{INT,{\rm H}\alpha}$ -- $\sigma_{gas,c}$ 
relation obtained for the  AGN-type. 
Inspection of panel b of Fig.~\ref{figure:r-r-int} shows that the location of the 
circumnuclear regions of the INT-type in the $L_{INT,{\rm H}\alpha}$ -- $\sigma_{gas,c}$ 
diagram, follow the trend outlined by the location of the AGN-type
for high values of the central gas velocity dispersion, and are slightly shifted toward  lower
luminosities for lower values of the central gas velocity dispersion. 

Panel c of Fig.~\ref{figure:r-r-int} shows the $L_{INT,OIII}$ -- $\sigma_{gas,c}$ diagram. 
Examination of this panel indicates that the $L_{INT,OIII}$ luminosities 
of the INT-type are slightly lower, on average, than 
those  of the AGN-type for similar values of the central gas velocity dispersion.

\subsection{Circumnuclear regions of the SF type}

\begin{figure}
\begin{center}
\resizebox{1.00\hsize}{!}{\includegraphics[angle=000]{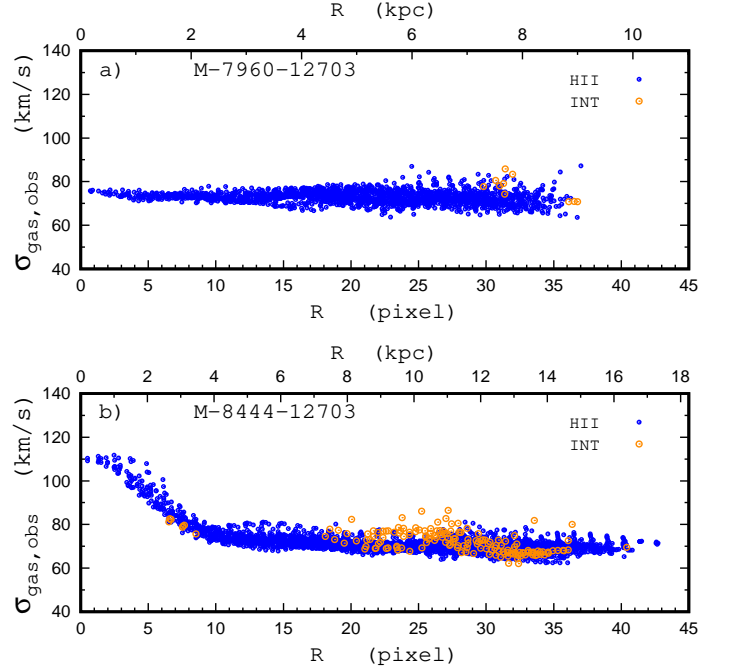}}
\caption{
  Gas velocity dispersion $\sigma_{gas}$ as a function of radius for two
  galaxies with the circumnuclear regions of the SF-type. 
  The spaxels of the SF-like spectra are denoted by  blue symbols, the spaxels 
  with  intermediate spectra are shown by  orange symbols.
}
\label{figure:s-r-sf}
\end{center}
\end{figure}

\begin{figure}
\begin{center}
\resizebox{0.85\hsize}{!}{\includegraphics[angle=000]{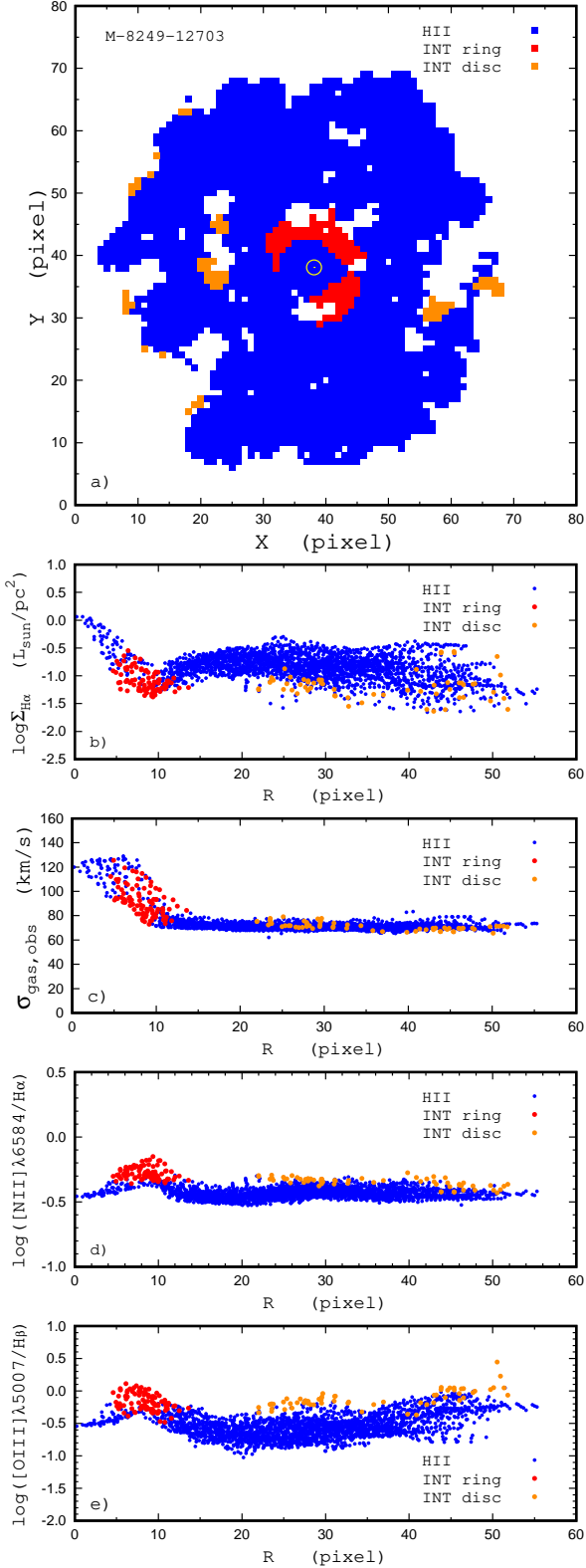}}
\caption{
Galaxy with a ring of the INT type radiation. 
{\em Panel} $a$: distribution of the spaxels with the SF-like (blue squares),
  intermediate spectra in the ring (red squares), and in the
disc (orange squares), for the galaxy
M-8249-12703. The circle marks the kinematical center of the galaxy.
{\em Panels} $b$ to $e$ show the variation with radius of 
the H$\alpha$ line surface brightness $\Sigma_{{\rm H}\alpha}$  ({\em panel} $b$),
the observed (non corrected fro the instrumental profile) gas velocity dispersion$\sigma_{gas,obs}$ ({\em panel} $c$), 
$X_{BPT}$ = log([N\,{\sc ii}]$\lambda$6584/H$\alpha$) ({\em panel} $d$), 
and $Y_{BPT}$ = log([O\,{\sc iii}]$\lambda$5007/H$\beta$) ({\em panel} $e$).
In each panel, 1 pixel corresponds to 0.276 kpc.
}
\label{figure:sf-ring-int}
\end{center}
\end{figure}

The H\,{\sc ii}-region-like  spectra are observed at the center of many 
MaNGA galaxies, this is, they are galaxies with the circumnuclear region of the SF-type. 
Fig.~\ref{figure:s-r-sf}  shows the observed (non corrected for the instrumental 
profile) gas velocity dispersion $\sigma_{gas,obs}$ as 
a function of radius for two galaxies with the circumnuclear regions of the SF-type. 
The spaxels with the SF-like spectra are denoted by blue symbols, and the spaxels 
with  the intermediate spectra in the disc  by  orange symbols.
From one side, the gas velocity dispersion $\sigma_{gas,obs}$ can be nearly 
constant over the whole galaxy, this is, the gas velocity dispersion $\sigma_{gas,obs}$ 
at the center of the circumnuclear region of the SF-type can be
close to that of the disc, see panel a of Fig.~\ref{figure:s-r-sf}. 
On the other hand, the gas velocity dispersion $\sigma_{gas,obs}$ at the center of 
the circumnuclear region of the SF-type can be significantly higher than that 
in the disc, see panel b of Fig.~\ref{figure:s-r-sf}. 

The central gas velocity dispersion $\sigma_{gas,c}$, the values of the central 
surface brightness in the H$\alpha$ emission line ($\Sigma_{{\rm H}\alpha,c}$) and 
 [O\,{\sc iii}]$\lambda\lambda$4959,5007 emission lines ($\Sigma_{OIII,c}$)  
in the circumnuclear regions of the SF-type are determined using the same 
approach as for the  AGN and INT-types. 
The mean value of the gas velocity dispersion in five spaxels with higher 
$\sigma_{gas}$ within the radius of 3 pixel is adopted as the central gas velocity 
dispersion  $\sigma_{gas,c}$. The central surface brightness in the H$\alpha$ 
and in the [O\,{\sc iii}]$\lambda\lambda$4959,5007 emission lines are  
determined as the mean values for the same five spaxels. 
 
In a several MaNGA galaxies (M-8137-09102, M-8249-12703, M-9500-12702), 
the circumnuclear region of the SF type is surrounded by a ring of  spaxels 
of  intermediate type spectra. 
For instance, panel $a$ of Fig.~\ref{figure:sf-ring-int} shows the distribution of the 
H\,{\sc ii}-region-like spaxels (blue symbols), 
the spaxels with the intermediate spectra in the circumnuclear ring (red symbols), 
and the spaxels with the intermediate spectra in the disc (orange symbols). 
The circle marks the kinematical center of  the galaxy.
Panels b, c, d, and e of  Fig.~\ref{figure:sf-ring-int} show the radial distributions of different 
characteristics (surface brightness in the H$\alpha$ emission line, gas velocity 
dispersion, and  BPT line ratios) for this galaxy. 

Two possible explanations of the appearance of the ring of the spaxels with
the spectra of the intermediate type around the innermost region of the SF
radiation can be suggested. First.  The spaxels with the spectra of the
intermediate type  lie at a dip in the H$\alpha$ surface brightness and do not
form a clear mixing line trajectory on the BPT diagram. This can indicate
that diffuse ionized gas might make a significant contribution to the line
emission like that in the spaxels with spectra of the INT type in the disc
far from the center.

Second. Such circumnuclear region looks as if the innermost AGN zone is replaced
by a SF-zone in the circumnuclear region of an AGN type.  
One can speculate that this configuration of the radiation distribution in the  
circumnuclear region in those galaxies can be explained in the following way. 
A star formation burst occurs at the center of a galaxy with the circumnuclear 
region of the AGN type.  
Then the contribution of the SF to the radiation becomes dominant at the center, 
moving its location on the BPT diagram from an AGN to an H\,{\sc ii}-region locus. 
The contribution of the SF to the radiation decreases with radius,  as seen in
panel $b$ of Fig.~\ref{figure:sf-ring-int}, and becomes low in
comparison to the contribution of the AGN (shock) at some radius, such that the spectra are
of the INT type in the ring. 
However, a detailed model should be constructed in order to confirm or reject 
this picture.

Thus, the central gas velocity dispersion can be either low or high in the circumnuclear region
of the SF-type. 
In  several galaxies, the circumnuclear region of the SF-type is surrounded by a ring of
 spaxels with the intermediate spectra.

\subsection{Detectability of the circumnuclear regions of the AGN type
from the optical spectra}

\begin{figure}
\begin{center}
\resizebox{1.00\hsize}{!}{\includegraphics[angle=000]{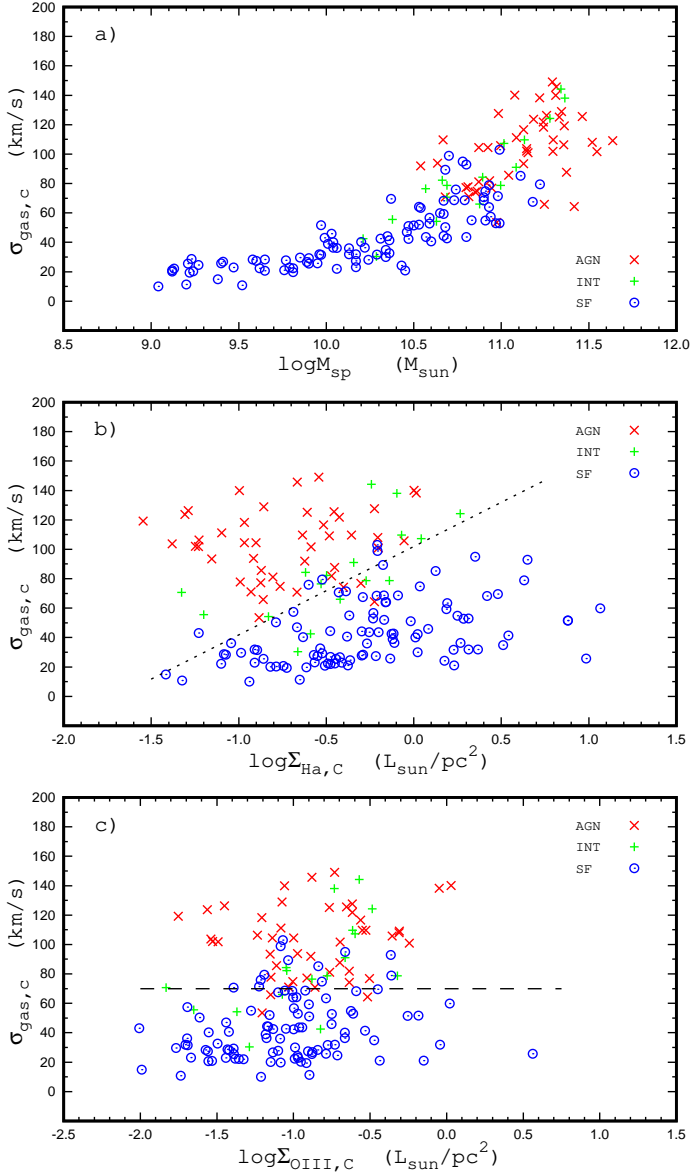}}
\caption{
Central gas velocity dispersion $\sigma_{gas,c}$ as a function of stellar mass of the galaxy $M_{sp}$  ({\em panel} $a$),
central surface brightness in the H$\alpha$ line $\Sigma_{{\rm H}\alpha,C}$ ({\em panel} $b$), 
and central surface brightness in the [O\,{\sc iii}]$\lambda\lambda$4959,5007 emission lines  $\Sigma_{OIII,C}$  ({\em panel} $c$).
Galaxies with the circumnuclear regions of the AGN-type are denoted by red crosses,
INT-type by green plus symbols, and SF-type by blue circles.
The line in panels b and c is the demarcation line between the location of  
galaxies with the circumnuclear regions of the AGN and SF-types.
}
\label{figure:sb-sv}
\end{center}
\end{figure}

We  argued above that the value of the central gas velocity dispersion $\sigma_{gas,c}$
can serve as an indicator of the AGN activity. 
Panel $a$ of Fig.~\ref{figure:sb-sv}  shows the central gas velocity dispersion 
as a function of the stellar mass for galaxies with the circumnuclear
regions of the AGN-type (red crosses),  INT-type (green plus symbols), and  
 SF-type (blue circles).  Inspection of panel $a$ of Fig.~\ref{figure:sb-sv}
shows that the central gas velocity dispersion increases significantly for galaxies with stellar masses
larger than $\sim$10$^{10.6}M_{\odot}$. Then one can expect an appreciable AGN activity
for those galaxies.

The values of the central gas velocity dispersion 
in  galaxies with the circumnuclear regions of the AGN-type are higher than those
for  galaxies with the circumnuclear regions of the SF-type. However, the  
diapason of  values of the central gas velocity dispersion in  galaxies of the AGN-type 
circumnuclear regions overlaps partially  with the values of the central gas velocity dispersion
in the galaxies with the circumnuclear regions of the SF-type. On one side, this overlapping can be attributed
to the errors in the determination of the central gas velocity dispersion (and/or mass
of the galaxy). On the other side, the lack of a sharp demarcation value in
the central gas velocity dispersions for galaxies with the circumnuclear regions of AGN and  SF-types  
can be considered as evidence of additional parameters playing a role.
If this is the case, then the presence of  AGN activity at a given level in the
galaxy does not necessary result in the appearance of the circumnuclear region of 
the AGN type in the BPT diagram.

Panel b of Fig.~\ref{figure:sb-sv}  shows the central gas velocity dispersion $\sigma_{gas,c}$ as
a function of the central surface brightness in the H$\alpha$ emission line $\Sigma_{{\rm H}\alpha,c}$ 
for  galaxies with the circumnuclear region of the AGN-type (red crosses) and for  galaxies
with the circumnuclear region of the SF-type (blue circles). A comparison between panels a and b of
Fig.~\ref{figure:sb-sv} shows that the demarcation line between the location of the AGN and  SF-types
in the  $\sigma_{gas,c}$ -- $\Sigma_{{\rm H}\alpha,c}$ diagram is more
clear than that in the $\sigma_{gas,c}$ -- M$_{sp}$ diagram.
The demarcation value between the central gas velocity dispersions in the circumnuclear regions of the AGN
and the SF-types depends on the value of the central H$\alpha$ surface brightness,
in the sense that the circumnuclear region at a given value of the central gas velocity dispersion
is detected as an AGN-type object in the BPT diagram  only if the central surface brightness in
 H$\alpha$  is lower than some value, or the circumnuclear region with a given value of
$\Sigma_{{\rm H}\alpha,c}$ will be detected in the BPT diagram as an AGN-type object  only if the
central gas velocity dispersion is higher than some value.

Panel c of Fig.~\ref{figure:sb-sv}  shows the central gas velocity dispersion $\sigma_{gas,c}$ as
a function of the central surface brightness in the [O\,{\sc iii}]$\lambda\lambda$4959,5007 emission
lines $\Sigma_{OIII,c}$ 
for the galaxies with the circumnuclear regions of the AGN-type (red crosses) and for  galaxies
with the circumnuclear regions of the SF-type (blue circles). Comparison between panels c and b of
Fig.~\ref{figure:sb-sv} shows that the demarcation line between the locations of the AGN and  SF-types
in the  $\sigma_{gas,c}$ -- $\Sigma_{OIII,C}$ diagram is less 
distinctive than that of the $\sigma_{gas,c}$ -- $\Sigma_{{\rm H}\alpha,c}$ diagram.

Thus, the AGN activity takes place at the center of each massive galaxy as it is indicated by the central gas
velocity dispersion.  The circumnuclear region is detected as an AGN-type object in the BPT diagram if 
the central gas velocity dispersion is higher than some value.
The demarcation value between the central gas velocity dispersion in the circumnuclear regions of the AGN
and the SF types can also depend on the value of the central H$\alpha$ surface brightness.

\section{Discussion}

The median value of the gas velocity dispersions in the spaxels with the  H\,{\sc ii}-region-like
spectra in the galaxy was used in the current study in the correction of the observed gas velocity
dispersions for the instrumental profile. Now we discuss this approach.
The mean values  of the gas velocity dispersion for  spaxels with H\,{\sc ii}-region-like spectra in the galaxy
can be slightly higher than the median values because of 
the gas velocity dispersions in some spaxels (close to the center of the galaxy)
can be higher than the average value for the disc. Then the median value of the gas velocity dispersion 
better specifies the gas velocity dispersion in the disc.

We have examined the median value of the gas velocity dispersions in the disc 
as a function of the galaxy inclination $i$, the distance to the galaxy $d$,
the stellar mass $M_{sp}$, and the central oxygen abundance in the galaxy 12+log(O/H). 
The median value for the 161 galaxies which is equal to 72.5 km/s, and 
the mean value of the scatter around the median value is 2.8 km/s.
The relative low scatter in the median values of the observed gas velocity dispersion in the discs  
of different galaxies is a strong evidence in favor  that the conditions of the observations,
as well as the physical conditions of the galaxies do not significantly influence on the median observed
gas velocity dispersion.
Indeed, the lack of  correlation between the median observed gas velocity dispersion and galaxy
inclination $i$ implies that the rotational velocity does not make an appreciable contribution
to the observed gas velocity dispersion.  The value of the observed gas velocity dispersion
does not depend on the  physical size of the area in the galaxy within the spaxel (which depends
on the distance to the galaxy $d$). The electron temperature in the   H\,{\sc ii} region
correlates with its oxygen abundance in the sense that the electron temperatures are higher
in the low-metallicity   H\,{\sc ii} regions. 
Then, the lack of the correlation between the observed gas velocity dispersion and oxygen
abundance implies that the electron temperature does not significantly influence  
the observed gas velocity dispersion.  At last, the observed gas velocity dispersion
is similar in galaxies of different masses.

The median value of the observed gas velocity dispersion in galaxies (72.5 $\pm$ 2.8 km/s)
is within the expected interval of the instrumental dispersion for the MaNGA survey,
50 -- 80 km/s \citep{Bundy2015}, and is close to the estimated instrumental dispersion
$\sim$70 km/s \citep{Westfall2019}.

Thus, the use of the median value of the gas velocity dispersions in the spaxels with
the  H\,{\sc ii}-region-like spectra in the galaxy as the instrumental profile is
quite justified (at least as a first-order approximation).

We have considered the circumnuclear regions of different BPT types. 
There are four configurations of the radiation distributions in the  circumnuclear regions in
(massive) galaxies: \\
1) AGN+INT, the innermost region of the AGN-like radiation is surrounded by  a ring of radiation
of  intermediate type, \\
2) INT, the spectra of the spaxels in the central region are of the intermediate type,  \\
3) SF+INT, the inner region of the   H\,{\sc ii}-like radiation is surrounded by  a ring of  
radiation of the intermediate type, \\
4) SF, the central area involves the spaxels with the  H\,{\sc ii}-like radiation only. \\

Figs.~\ref{figure:bpt-definition}, ~\ref{figure:rg-param-example}, ~\ref{figure:int-example},
and ~\ref{figure:sf-ring-int}  illustrate those configurations of the radiation distributions in the  
circumnuclear regions. 
Fig.~\ref{figure:bpt-definition} and Fig.~\ref{figure:rg-param-example} show the galaxy M-7495-12704
which is an example of galaxies with circumnuclear regions of the AGN+INT configuration. 
Fig.~\ref{figure:int-example} shows the galaxy M-8137-12703, an example of galaxies with the circumnuclear 
regions of the INT configuration. 
Fig.~\ref{figure:sf-ring-int} shows the galaxy M-8249-12703, an example of galaxies with the circumnuclear 
regions of the SF+INT configuration. 
The sequence of those configurations (from the AGN+INT through the 
INT (or SF+INT) to the SF) can be explained by an increase of the contribution of the SF 
to the radiation in the circumnuclear region. 
 
Thus, we suggest that the gas velocity dispersion can serve as an indicator of  AGN activity.
Fig.~\ref{figure:sb-sv}  shows that the value of central gas velocity dispersion depends on stellar mass.
This correlation can be explained if gas velocity dispersion is a tracer of gaseous bulge (related to
the host galaxy mass).
However, this correlation can be explained equally well if gas velocity dispersion is a tracer of
AGN activity (related to the black hole mass) since a correlation exists between the mass of the
black hole and stellar mass of the host galaxy.
Our suggestion that the gas velocity dispersion can serve as an indicator of  AGN activity is based
on the fact that the characteristics of the gas velocity dispersion correlate with 
direct characteristics of the AGN activity.
The radius of the zone with enhanced gas velocity dispersion is related to the radius of the
zone of the influence of AGN.
The values of the central gas velocity dispersion correlates with AGN luminosity 
in [O\,{\sc iii}]$\lambda\lambda$4959,5007 and H$\alpha$  emission lines.
We did not find facts which are in conflict with our suggestion.

\section{Conclusions}

We have considered the circumnuclear regions in star-forming MaNGA galaxies.
The spaxels spectra are classified as AGN-like,  H\,{\sc ii}-region-like (or SF-like), and 
intermediate (INT) spectra  according to their positions in the Baldwin-Phillips-Terlevich (BPT) diagram. 
We found that there are four configurations of the radiation distributions in the  circumnuclear regions in
galaxies: \\
1) AGN+INT, the innermost region of the AGN-like radiation is surrounded by a ring of radiation
of the intermediate type, \\
2) INT, the radiation at the center of galaxy is the intermediate type, \\
3) SF+INT, the inner region of the   H\,{\sc ii}-region-like radiation is surrounded by a ring of 
radiation of the intermediate type, \\
4) SF, the central area involves the spaxels with the  H\,{\sc ii}-region-like radiation only. \\
The sequence of those configurations (from the AGN+INT through the INT (or SF+INT) to the SF) can be explained
by an increase in the contribution of the SF to the radiation of the circumnuclear region. 

The low ionization nuclear emission line regions (LINERs) of configurations 1 and 2 are examined.  
The spaxel spectra in the LINERs form a sequences on the BPT diagram,
that is, they lie along the known AGN-SF mixing line trajectories.  
The diagnostic line ratios in the spaxels spectra change smoothly with radius, 
from AGN-like (or INT)  in the spaxels spectra in the galactic center, to 
H\,{\sc ii}-region-like  in the spaxels spectra at larger galactocentric distances. 
This evidences  that the change of the diagnostic line ratio along radius
is primary caused by variations in the fraction of the line emission excited by AGN activity. 
Our results are in agreement with the widely accepted paradigm that the LINERs are excited by 
AGN activity  and also agree with the results from \citet{Davies2014a,Davies2014b,Davies2016}.  

The AGN and INT radiation in the circumnuclear region is accompanied by an enhancement in the 
gas velocity dispersion $\sigma_{gas}$. The radius of the area of the AGN and INT radiation
(radius of influence of the AGN on the radiation) $R_{INT}$ 
is similar to the radius of the area with enhanced gas velocity dispersion $R_{\sigma_{gas}}$,
and the central gas velocity dispersion $\sigma_{gas,c}$ correlates with the luminosity of the AGN+INT area. 
This allows us to assume that  the central gas velocity dispersion can serve as indicator of  AGN activity.

An appreciable  $\sigma_{gas,c}$ were also measured in the SF-type centers of massive galaxies.
With our assumption  that  the central gas velocity dispersion can serve as indicator of
 AGN activity, this implies the AGN activity in those galaxies. The interval 
of the values of the central gas velocity dispersion in the galaxies with the circumnuclear
regions of the AGN type overlaps partly with the interval of those in the  galaxies with
the circumnuclear regions of the SF type. The lack of the sharp demarcation between the values 
of the central gas velocity dispersions in the circumnuclear regions of the AGN and the SF
types can be considered as an evidence in favor of that the manifestation of the circumnuclear
region as AGN or as SF depends not only on the value of the central gas velocity dispersion
(the level of the AGN activity) but it is also governed by an additional parametr(s). 
We found that there is a demarcation line between the positions of the AGN-type and SF-type galaxies
on the central surface brightness in the H$\alpha$ line -- central gas velocity dispersion diagram
in the sense that the galaxies with a given value of the central gas velocity dispersion are 
detected as the AGN-type object on the BPT diagram if only the central surface brightness in the
H$\alpha$ line is lower than some value. 
 One can assume that the value of the central gas velocity dispersion may be the more
sensitive indicator of the AGN activity than the position on the BPT diagram.

\section*{Acknowledgements}

We are grateful to the referee for his/her constructive comments. \\
L.S.P., E.K.G., and I.A.Z.\  acknowledge support within the framework
of Sonderforschungsbereich (SFB 881) on ``The Milky Way System''
(especially subproject A5), which is funded by the German Research
Foundation (DFG). \\ 
L.S.P.\ and I.A.Z.\ thank for hospitality of the
Astronomisches Rechen-Institut at Heidelberg University, where part of
this investigation was carried out. \\
L.S.P acknowledges support within the framework of the program of the NAS of
Ukraine “Support for the development of priority fields of scientific
research” (CPCEL 6541230)" \\
I.A.Z acknowledges support by the grant for young scientist’s research
laboratories of the National Academy of Sciences of Ukraine. \\
M.A.L.-L. is a DARK-Carlsberg Foundation Fellow (Semper Ardens project CF15-0384) \\
The work is performed according to the Russian Government Program of Competitive Growth
of Kazan Federal University and Russian Science Foundation, grant no. 20-12-00105 \\
We acknowledge the usage of the HyperLeda database (http://leda.univ-lyon1.fr). \\
Funding for SDSS-III has been provided by the Alfred P. Sloan Foundation,
the Participating Institutions, the National Science Foundation,
and the U.S. Department of Energy Office of Science.
The SDSS-III web site is http://www.sdss3.org/. \\
Funding for the Sloan Digital Sky Survey IV has been provided by the
Alfred P. Sloan Foundation, the U.S. Department of Energy Office of Science,
and the Participating Institutions. SDSS-IV acknowledges
support and resources from the Center for High-Performance Computing at
the University of Utah. The SDSS web site is www.sdss.org. \\
SDSS-IV is managed by the Astrophysical Research Consortium for the 
Participating Institutions of the SDSS Collaboration including the 
Brazilian Participation Group, the Carnegie Institution for Science, 
Carnegie Mellon University, the Chilean Participation Group,
the French Participation Group, Harvard-Smithsonian Center for Astrophysics, 
Instituto de Astrof\'isica de Canarias, The Johns Hopkins University, 
Kavli Institute for the Physics and Mathematics of the Universe (IPMU) / 
University of Tokyo, Lawrence Berkeley National Laboratory, 
Leibniz Institut f\"ur Astrophysik Potsdam (AIP),  
Max-Planck-Institut f\"ur Astronomie (MPIA Heidelberg), 
Max-Planck-Institut f\"ur Astrophysik (MPA Garching), 
Max-Planck-Institut f\"ur Extraterrestrische Physik (MPE), 
National Astronomical Observatories of China, New Mexico State University, 
New York University, University of Notre Dame, 
Observat\'ario Nacional / MCTI, The Ohio State University, 
Pennsylvania State University, Shanghai Astronomical Observatory, 
United Kingdom Participation Group,
Universidad Nacional Aut\'onoma de M\'exico, University of Arizona, 
University of Colorado Boulder, University of Oxford, University of Portsmouth, 
University of Utah, University of Virginia, University of Washington, University of Wisconsin, 
Vanderbilt University, and Yale University.

\end{document}